\documentclass[leqno,11pt]{article}
\usepackage{amsfonts,latexsym}
\usepackage{amsmath}
\usepackage{amscd}
\usepackage{float,amsmath,amssymb,mathrsfs,bm,multirow,graphics}
\usepackage[dvips]{graphicx}

\newcommand{\nequation}{\setcounter{equation}{0}}
\renewcommand{\theequation}{\mbox{\arabic{section}.\arabic{equation}}}
\newcommand{\R}{{\Bbb R}}

\newcommand{\C}{{\Bbb C}}
\newcommand{\Q}{{\Bbb Q}}
\newcommand{\proofbegin}{\noindent{\it Proof\,\,}}
\newcommand{\proofend}{$\hfill\Box$\bigskip}

\newtheorem{theorem}{Theorem}[section]
\newtheorem{proposition}[theorem]{Proposition}
\newtheorem{lemma}[theorem]{Lemma}
\newtheorem{definition}[theorem]{Definition}
\newtheorem{assumption}[theorem]{Assumption}

\newtheorem{figuretext}{Figure}

\input epsf
\title{\sc The derivative nonlinear Schr\"odinger equation on the half-line}
\author{Jonatan Lenells}
\date{{\small Department of Applied Mathematics and Theoretical Physics, University of Cambridge, Cambridge CB3 0WA, United Kingdom}}

\begin{document}
\maketitle

\begin{abstract} 
\noindent
We analyze the derivative nonlinear Schr\"odinger equation $iq_t + q_{xx} = i\left(|q|^2q\right)_x$ on the half-line using the Fokas method. Assuming that the solution $q(x,t)$ exists, we show that it can be represented in terms of the solution of a matrix Riemann-Hilbert problem formulated in the plane of the complex spectral parameter $\zeta$. The jump matrix has explicit $x,t$ dependence and is given in terms of the spectral functions $a(\zeta)$, $b(\zeta)$ (obtained from the initial data $q_0(x) = q(x,0)$) as well as $A(\zeta)$, $B(\zeta)$ (obtained from the boundary values $g_0(t) = q(0,t)$ and $g_1(t) = q_x(0,t)$). The spectral functions are not independent, but related by a compatibility condition, the so-called global relation. Given initial and boundary values $\{q_0(x), g_0(t), g_1(t)\}$ such that there exist spectral function satisfying the global relation, we show that the function $q(x,t)$ defined by the above Riemann-Hilbert problem exists globally and solves the derivative nonlinear Schr\"odinger equation with the prescribed initial and boundary values. 
\end{abstract}

\noindent
{\small{\sc PACS numbers (2008)}: 02.30.Ik, 42.81.Dp, 52.35.Bj.}


\section{Introduction}\nequation
The derivative nonlinear Schr\"odinger (DNLS) equation
\begin{equation}\label{DNLS} 
  iq_t + q_{xx} = i\left(|q|^2q\right)_x,
\end{equation}
has several applications in e.g. plasma physics and nonlinear fiber optics. In plasma physics, it is a model for Alfv\'en waves propagating parallel to the ambient magnetic field, $q$ being the transverse magnetic field perturbation and $x$ and $t$ being space and time coordinates, respectively \cite{Mjolhus76}. In the context of fiber optics, equation (\ref{DNLS}) arises as follows. While the propagation of nonlinear pulses in optical fibers is described to first order by the nonlinear Schr\"odinger (NLS) equation, it is necessary when considering very short input pulses to include higher-order nonlinear effects \cite{Kodama1985}. When the effect of self-steepening $(s \neq 0)$ is included, the fundamental equation is
\begin{equation}\label{fiber}  
  iu_\xi + \frac{1}{2}u_{\tau\tau} + |u|^2u + i s\left(|u|^2u\right)_\tau  = 0,
\end{equation}
where $u$ is the amplitude of the complex field envelope, $\tau$ is a time variable, and $\xi$ measures the distance along the fiber with respect to a frame of reference moving with the pulse at the group velocity cf. \cite{Agrawal2007}. Equation (\ref{fiber}) is related to equation (\ref{DNLS}) by the change of variables
$$u(\xi,\tau) = q(x,t) e^{i\left(\frac{t}{4s^4}  - \frac{x}{2s^2}\right)}, \qquad \xi = \frac{t}{2s^2}, \qquad \tau = -\frac{x}{2s} + \frac{t}{2s^3}.$$

Being integrable, equation (\ref{DNLS}) admits an infinite number of conservation laws and can be analyzed by means of inverse scattering techniques both in the case of vanishing \cite{K-N} as well as nonvanishing \cite{K-I} boundary conditions. The soliton solutions have been investigated from several points of view (see e.g. \cite{IKWS, Lashkin}) and a tri-Hamiltonian structure was put forward in \cite{M-Z}.  

In this paper we use the general method for solving boundary value problems for nonlinear integrable PDEs announced in \cite{F1997} to study equation (\ref{DNLS}) on the half-line. 
Assuming that the solution $q(x,t)$ exists, we show that it can be represented in terms of the solution of a matrix Riemann-Hilbert (RH) problem formulated in the plane of the complex spectral parameter $\zeta$, with jump matrices given in terms of spectral functions $a(\zeta)$, $b(\zeta)$ (obtained from the initial data $q_0(x) = q(x,0)$) and $A(\zeta)$, $B(\zeta)$ (obtained from the boundary values $g_0(t) = q(0,t)$ and $g_1(t) = q_x(0,t)$). 

An important advantage of the methodology of \cite{F1997} is that it yields precise information about the long time asymptotic of the solution. Indeed, using the nonlinearization of the steepest descent method presented in \cite{D-Z}, it is possible to describe how the solution for large $t$ splits into a collection of solitons traveling at constant speeds of order $1$, while away from these solitons the asymptotics displays a dispersive character (see \cite{F-I-S} for a detailed description in the case of the NLS equation). The usefulness of the asymptotic information for a physical problem can be understood by considering an example where the field is at rest at some initial time $t = 0$. Assuming that we can (with the help of some kind of apparatus) create or measure the waves emanating from some fixed point, say $x = 0$, in space, we arrive at an initial-boundary value problem for which the initial data $q(x, 0)$ vanishes identically, while the boundary values $q(0,t)$ and $q_x(0,t)$ are at our disposal. An analysis of the asymptotic behavior of the solution then provides information about the long time effect of the known boundary values. In the context of water waves modeled by the KdV equation with $q(0,t)$ a given periodic function of $t$, this problem is addressed in \cite{B-F}. It is expected that similar considerations are valid for Alfv\'en plasma waves which are governed by the derivative NLS equation studied in this paper.

The fact that the spectral functions satisfy a so-called global relation imposes a constraint on the given initial and boundary values $\{q_0(x), g_0(t), g_1(t)\}$, so that not all of them can be specified independently. Whereas the problem of identifying the spectral functions corresponding to the unknown boundary values in the case of linear PDEs can be solved by means of algebraic manipulations, in the case of nonlinear PDEs this is in general a difficult and nonlinear problem. We do not analyze this issue in the case of equation (\ref{DNLS}) in this paper.\footnote{Even when one assumes vanishing initial data, $q_0(x) = 0$, the corresponding analysis for the NLS equation (whose Lax pair has a much simpler $t$-part) is highly nontrivial cf. \cite{F-I-S}.} Let us also mention that while the NLS equation admits a particular class of so-called linearizable boundary conditions for which this problem can be solved algebraically by symmetry considerations, we have not been able to identify any such boundary conditions in the case of equation (\ref{DNLS}). 
However, for boundary values such that the global relation is fulfilled, we show that the function $q(x,t)$ defined by the above Riemann-Hilbert problem exists globally and solves the DNLS equation (\ref{DNLS}) with the prescribed initial and boundary values.

This method has previously been applied to the nonlinear Schr\"odinger equation on the half-line \cite{F-I-S}. Although the analysis of (\ref{DNLS}) is in many ways similar to that of the NLS equation, it also presents some distinctive features: (i) In order to arrive at a Riemann-Hilbert problem with the appropriate boundary condition at infinity, the Lax pair first has to be transformed by introduction of a differential one-form $\Delta$. This modification is made possible since one of the conservation laws ascertains that $\Delta$ is closed. (ii) Since the Lax pair includes the spectral parameter $\zeta$ raised to the fourth power, the Riemann-Hilbert problem splits the complex $\zeta$-plane into eight different sectors rather than into four as is the case for NLS. Nevertheless, additional symmetry implies that only four different jump matrices have to be specified. (iii) Due to the presence of $\Delta$ the recovery of $q(x,t)$ from the solution of the Riemann-Hilbert problem is more involved.

Let us also point out that whereas the NLS equation comes in a focusing and a defocusing version, this distinction does not apply to (\ref{DNLS}) as the two equations
$$iq_t + q_{xx} = \pm i(|q|^2q)_x,$$
can be transformed into each other by replacing $x \to -x$.

Section \ref{spectralanalysissec} contains the spectral analysis of the Lax pair for (\ref{DNLS}). The spectral functions $a,b,A,B$ are further investigated in section \ref{spectralfunctionssec} with some proofs postponed to the two appendices. Finally, the Riemann-Hilbert problem is presented in section \ref{RHsec}. 

\section{Spectral analysis}\label{spectralanalysissec}\nequation
Equation (\ref{DNLS}) admits the Lax pair formulation \cite{K-N}
\begin{equation}\label{KNlaxpair}
\begin{cases}
	& v_{1x} + i\zeta^2 v_1 = q\zeta v_2,	\\
	& v_{2x} - i\zeta^2 v_2 = r\zeta v_1,
\end{cases} \qquad\qquad \begin{cases}
	& iv_{1t} = Av_1 + B v_2,	\\
	& iv_{2t} = Cv_1 - Av_2,
\end{cases}
\end{equation}
where $\zeta \in \C$ is the spectral parameter, $ r =  \bar{q}$, and
$$A= 2\zeta^4 + \zeta^2 rq, \qquad B= 2i\zeta^3q - \zeta q_x + i\zeta r q^2, \qquad C= 2i\zeta^3r + \zeta r_x + i\zeta r^2 q.$$
Letting 
$$\psi =  \begin{pmatrix} v_1 \\ v_2 \end{pmatrix}, \qquad Q = \begin{pmatrix} 0 & q \\ r & 0 \end{pmatrix}, \qquad \sigma_3 = \begin{pmatrix} 1 & 0 \\ 0 & -1 \end{pmatrix},$$
we can write (\ref{KNlaxpair}) as
\begin{equation}\label{psilax}
\begin{cases}
	& \psi_x + i\zeta^2\sigma_3 \psi = \zeta Q \psi, \\
	& \psi_t + 2i\zeta^4\sigma_3 \psi = \left(-i\zeta^2Q^2\sigma_3+ 2\zeta^3 Q - i\zeta Q_x\sigma_3 + \zeta Q^3\right)\psi.
\end{cases}
\end{equation}
Extending the column vector $\psi$ to a $2\times 2$ matrix and letting $\psi = \Psi e^{-i(\zeta^2 x + 2\zeta^4 t)\sigma_3}$, we obtain the equivalent Lax pair
\begin{equation}\label{Psilax}
\begin{cases}
	& \Psi_x + i\zeta^2[\sigma_3, \Psi] = \zeta Q \Psi, \\
	& \Psi_t + 2i\zeta^4[\sigma_3, \Psi] = (-i\zeta^2Q^2\sigma_3 + 2\zeta^3 Q - i\zeta Q_x\sigma_3 + \zeta Q^3) \Psi,
\end{cases}
\end{equation}
which can be written as
\begin{equation}\label{Psilaxdiffform}
d\left(e^{i(\zeta^2 x + 2\zeta^4 t)\hat{\sigma}_3} \Psi(x,t,\zeta) \right) = e^{i(\zeta^2 x + 2\zeta^4 t)\hat{\sigma}_3} U(x,t,\zeta)\Psi,
\end{equation}
where 
\begin{equation}\label{Udef}  
  U = U_1dx + U_2 dt =  \zeta Q dx+ \left(-i\zeta^2Q^2\sigma_3 + 2\zeta^3 Q - i\zeta Q_x\sigma_3 + \zeta Q^3\right)dt,
\end{equation}
and $\hat{\sigma}_3$ acts on a $2\times 2$ matrix $A$ by $\hat{\sigma}_3A = \sigma_3 A \sigma_3^{-1}$.
In order to formulate a Riemann-Hilbert problem for the solution of the inverse spectral problem, we seek solutions of the spectral problem which approach the $2 \times 2$ identity matrix $I$ as $\zeta \to \infty$. It turns out that solutions of equation (\ref{Psilaxdiffform}) do not exhibit this property; hence our next step is to investigate how equation (\ref{Psilaxdiffform}) for $\Psi$ can be transformed into an equation with the desired asymptotic behavior.

\subsection{Asymptotic behavior}\label{asymptoticsubsec}
Consider a solution of (\ref{Psilaxdiffform}) of the form
$$\Psi = D + \frac{\Psi_1}{\zeta} + \frac{\Psi_2}{\zeta^2} + \frac{\Psi_3}{\zeta^3} + O\left(\frac{1}{\zeta^4}\right), \quad \zeta \to \infty,$$
where $D, \Psi_1, \Psi_2, \Psi_3$ are independent of $\zeta$. Substituting the above expansion into the $x$-part of (\ref{Psilax}), it follows from the $O(\zeta^2)$ terms that $D$ is a diagonal matrix. Furthermore, one finds the following equations for the $O(\zeta)$ and the diagonal part of the $O(1)$ terms
$$O(\zeta): i[\sigma_3, \Psi_1] = QD, \quad \text{i.e.} \quad \Psi_1^{(o)} = \frac{i}{2}QD\sigma_3,$$
where $\Psi_1^{(o)}$ denotes the off-diagonal part of $\Psi_1$;
$$O(1): D_x = Q\Psi_1^{(o)},$$
i.e.
\begin{equation}\label{Dxeq}
  D_x = \frac{i}{2}Q^2 \sigma_3 D.
\end{equation}

On the other hand, substituting the above expansion into the $t$-part of (\ref{Psilax}), one obtains for the $O(\zeta^3)$ terms
\begin{equation}\label{Ozeta3terms}
O(\zeta^3): 2i[\sigma_3, \Psi_1] = 2QD, \quad \text{i.e.} \quad \Psi_1^{(o)} = \frac{i}{2}QD\sigma_3;
\end{equation}
for the off-diagonal part of the $O(\zeta)$ terms
\begin{equation}\label{orderonetpart}
  O(\zeta): 2i[\sigma_3, \Psi_3] = -iQ^2 \sigma_3\Psi_1^{(o)} + 2Q \Psi_2^{(d)} - iQ_x\sigma_3D + Q^3 D,
\end{equation}
i.e.
\begin{equation}\label{minusiQ2}
- iQ^2\sigma_3 \Psi_2^{(d)} + 2Q\psi_3^{(o)} = -\frac{1}{2}Q^3\Psi_1^{(o)} + \frac{1}{2}QQ_x D + \frac{i}{2}Q^4 \sigma_3 D,
\end{equation}
where $\Psi_2^{(d)}$ denotes the diagonal part of $\Psi_2$; and for the diagonal part of the $O(1)$ terms
$$O(1): D_t =  - iQ^2\sigma_3 \Psi_2^{(d)} + 2Q\psi_3^{(o)} - iQ_x\sigma_3 \Psi_1^{(o)} + Q^3 \Psi_1^{(o)},$$
i.e., using (\ref{Ozeta3terms}) and (\ref{minusiQ2}),
$$
D_t = \left(\frac{3i}{4} Q^4 \sigma_3  + \frac{1}{2}[Q, Q_x]\right)D.
$$
We can write this in terms of $q$ and $r$ as
\begin{equation}\label{Dteq}
  D_t = \left(\frac{3i}{4} r^2q^2 + \frac{1}{2}(r_xq - rq_x)\right)\sigma_3 D.
\end{equation}

\subsection{Transformed Lax pair} 
Equation (\ref{DNLS}) admits the conservation law
$$\left(\frac{i}{2}rq\right)_t - \left(\frac{3i}{4} r^2q^2 + \frac{1}{2}(r_xq - rq_x)\right)_x = 0.$$
Consequently, the two equations (\ref{Dxeq}) and (\ref{Dteq}) for $D$ are consistent and are both satisfied if we define
\begin{equation}\label{Ddef}  
  D(x,t) = e^{i\int^{(x,t)}_{(\infty, 0)} \Delta \sigma_3},
\end{equation}
where $\Delta$ is the closed real-valued one-form
\begin{equation}\label{Deltadef}  
  \Delta(x,t) = \frac{1}{2}rq dx + \left(\frac{3}{4} r^2q^2 - \frac{i}{2}(r_xq - rq_x)\right)dt.
\end{equation}
We note that the integral in (\ref{Ddef}) is independent of the path of integration and that $\Delta$ is independent of $\zeta$.

This asymptotic analysis suggests that we introduce a new function $\mu$ by $\Psi = \mu D$. Actually it is convenient to introduce $\mu$ via
\begin{equation}\label{Psimurelation}
\Psi(x,t,\zeta) = e^{i\int^{(x,t)}_{(0, 0)} \Delta \hat{\sigma}_3}\mu(x,t,\zeta) D(x,t).
\end{equation}
Then 
\begin{equation}\label{muasymptotic}  
  \mu = I + O\left(\frac{1}{\zeta}\right), \qquad \zeta \to \infty,
\end{equation}
and the Lax pair (\ref{Psilaxdiffform}) becomes
\begin{equation}\label{laxdiffform}  
  d\left(e^{i(\zeta^2 x + 2\zeta^4 t)\hat{\sigma}_3} \mu(x,t,\zeta) \right) = W(x,t,\zeta),
\end{equation}
where $W(x,t,\zeta) = e^{i(\zeta^2 x + 2\zeta^4 t)\hat{\sigma}_3} V(x,t,\zeta)\mu(x,t,\zeta)$ and
\begin{align*}
V =& V_1dx + V_2 dt = e^{-i\int^{(x,t)}_{(0, 0)} \Delta \hat{\sigma}_3}(U -i\Delta \sigma_3).
\end{align*}
Taking into account the definitions of $U$ and $\Delta$, we find
\begin{align}\label{V1explicit}
V_1 =& \begin{pmatrix} -\frac{i}{2} rq	&	\zeta q e^{-2i\int^{(x,t)}_{(0, 0)} \Delta} \\
\zeta r e^{2i\int^{(x,t)}_{(0, 0)} \Delta}	&	\frac{i}{2} r q \end{pmatrix},
	\\	\label{V2explicit}
V_2 =& \begin{pmatrix} -i\zeta^2rq - \frac{3i}{4}r^2q^2 - \frac{1}{2}(r_xq - rq_x)		&	\left(2\zeta^3 q + i\zeta q_x + \zeta q^2 r\right)e^{-2i\int^{(x,t)}_{(0, 0)} \Delta}	\\
\left(2\zeta^3 r - i\zeta r_x + \zeta r^2 q\right)e^{2i\int^{(x,t)}_{(0, 0)} \Delta}	&	i\zeta^2rq + \frac{3i}{4}r^2q^2 + \frac{1}{2}(r_xq - rq_x)	\end{pmatrix},
\end{align}
and equation (\ref{laxdiffform}) for $\mu$ can be written as
\begin{equation}\label{mulax}  
\begin{cases}
	& \mu_x + i\zeta^2 [\sigma_3, \mu] = V_1\mu, \\
	& \mu_t + 2i\zeta^4 [\sigma_3, \mu] = V_2\mu.
\end{cases}
\end{equation}

\subsection{Bounded and analytic eigenfunctions}
\begin{figure}
\begin{center}
    \includegraphics[width=.3\textwidth]{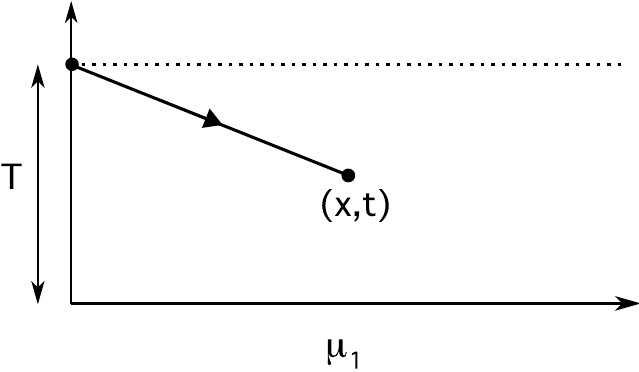} \quad
    \includegraphics[width=.3\textwidth]{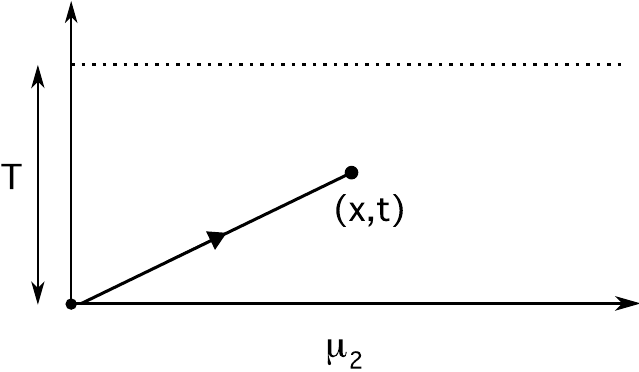} \quad
    \includegraphics[width=.3\textwidth]{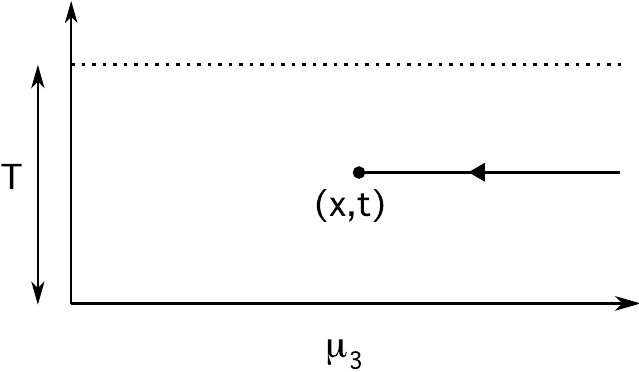} \\
     \begin{figuretext}\label{mucontours.pdf}
       The solutions $\mu_1$, $\mu_2$, and $\mu_3$ of (\ref{laxdiffform}).
     \end{figuretext}
     \end{center}
\end{figure}

Let equation (\ref{laxdiffform}) be valid for $x$ and $t$ in 
$$\Omega = \{0 < x < \infty, \ 0 < t< T\},$$
where $T \leq \infty$ is a given positive constant; unless otherwise specified, we suppose that $T < \infty$.
Assume that the function $q(x,t)$ has sufficient smoothness and decay.
Define three solutions $\mu_j$, $j = 1,2,3$, of (\ref{laxdiffform}) by
\begin{equation}\label{mujdef}  
  \mu_j(x,t,\zeta) = I + \int_{(x_j, t_j)}^{(x,t)} e^{-i(\zeta^2 x + 2\zeta^4 t)\hat{\sigma}_3}W(x',t',\zeta),
\end{equation}
where $(x_1, t_1) = (0, T)$, $(x_2, t_2) = (0, 0)$, and $(x_3, t_3) = (\infty, t)$. Since the one-form $W$ is exact, the integral on the right-hand side is independent of the path of integration. We choose the particular contours shown in Figure \ref{mucontours.pdf}. This choice implies the following inequalities on the contours,
\begin{align*}
(x_1, t_1) \to (x,t): x' - x \leq 0,& \qquad t' - t \geq 0,
	\\
(x_2, t_2) \to (x,t): x' - x \leq 0,& \qquad t' - t \leq 0,
	\\
(x_3, t_3) \to (x,t): x' - x \geq 0.&
\end{align*}
The second column of the matrix equation (\ref{mujdef}) involves $\exp[2i(\zeta^2(x' - x) + 2\zeta^4(t' - t))]$. Using the above inequalities it follows that this exponential is bounded in the following regions of the complex $\zeta$-plane,
\begin{align*}
(x_1, t_1) \to (x,t): \{\text{\upshape Im}\, \zeta^2 \leq 0\} &\cap \{\text{\upshape Im}\, \zeta^4 \geq 0\},
	\\
(x_2, t_2) \to (x,t): \{\text{\upshape Im}\, \zeta^2 \leq 0\} &\cap \{\text{\upshape Im}\, \zeta^4 \leq 0\},
	\\
(x_3, t_3) \to (x,t): \{\text{\upshape Im}\, \zeta^2 \geq 0\}&.
\end{align*}
Thus the second column vectors of $\mu_1$, $\mu_2$, and $\mu_3$ are bounded and analytic for $\zeta \in \C$ such that $\zeta^2$ belongs to the third quadrant, fourth quadrant, and the upper half-plane, respectively. We will denote these vectors with superscripts $(3)$, $(4)$, and $(12)$ to indicate these boundedness properties.
Similar conditions are valid for the first column vectors, and so
$$\mu_1 = \left(\mu_1^{(2)}, \mu_1^{(3)}\right), \qquad 
\mu_2 = \left(\mu_2^{(1)}, \mu_2^{(4)}\right), \qquad 
\mu_3 = \left(\mu_3^{(34)}, \mu_3^{(12)}\right).$$
We note that $\mu_1$ and $\mu_2$ are entire functions of $\zeta$. By (\ref{muasymptotic}), it holds that
$$\mu_j(x, t, \zeta) = I + O\left(\frac{1}{\zeta}\right), \qquad \zeta \to \infty, \quad j = 1,2,3.$$
The $\mu_j$'s are the fundamental eigenfunctions needed for the formulation of a Riemann-Hilbert problem in the complex $\zeta$-plane. Indeed, in each region 
$$D_j = \left\{\zeta \in \C | \arg \zeta^2 \in \left((j-1)\pi/2, j\pi/2\right)\right\}, \qquad j = 1, \dots, 4,$$
of the complex $\zeta$-plane (see Figure \ref{xtRHproblem.pdf}), there exist two column vectors which are bounded and analytic. For example, in $D_1$ these two vectors are $\mu_2^{(1)}$ and $\mu_3^{(12)}$. 

In order to derive a Riemann-Hilbert problem, we only have to compute the `jumps' of these vectors across the boundaries of the $D_j$'s.
It turns out that the relevant jump matrices can be uniquely defined in terms of two $2\times 2$-matrix valued spectral functions $s(\zeta)$ and $S(\zeta)$ defined as follows. 
Any two solutions $\mu$ and $\tilde{\mu}$ of (\ref{laxdiffform}) are related by an equation of the form
\begin{equation}\label{mutildemu}  
  \mu(x,t,\zeta) = \tilde{\mu}(x,t,\zeta)  e^{-i(\zeta^2 x + 2\zeta^4 t)\hat{\sigma}_3} C_0(\zeta),
\end{equation}
where $C_0(\zeta)$ is a $2\times 2$ matrix independent of $x$ and $t$. Indeed, let $\psi$ and $\tilde{\psi}$ be the solutions of equation (\ref{psilax}) corresponding to $\mu$ and $\tilde{\mu}$ according to
\begin{equation}\label{psimurelation}
\psi(x,t,\zeta) = e^{i\int^{(x,t)}_{(0, 0)} \Delta \hat{\sigma}_3}\mu(x,t,\zeta) D(x,t)e^{-i(\zeta^2 x + 2\zeta^4 t) \sigma_3}.
\end{equation}
Then, since the first and second columns of a solution of (\ref{psilax}) satisfy the same equation, there exists a $2 \times 2$ matrix $C_1(\zeta)$ independent of $x$ and $t$ such that
\begin{equation}\label{psirelations}  
  \psi(x,t, \zeta) = \tilde{\psi}(x,t, \zeta) C_1(\zeta).
\end{equation}
It follows that (\ref{mutildemu}) is satisfied with $C_0(\zeta) = e^{-i\int^{(\infty, 0)}_{(0, 0)} \Delta \hat{\sigma}_3} C_1(\zeta)$. 
We define $s(\zeta)$ and $S(\zeta)$ by the relations
\begin{align}
\label{seq} 
  \mu_3(x,t,\zeta) &= \mu_2(x,t,\zeta)e^{-i(\zeta^2 x + 2\zeta^4 t)\hat{\sigma}_3} s(\zeta),
		\\
  \label{Seq} 
  \mu_1(x,t,\zeta) &= \mu_2(x,t,\zeta)  e^{-i(\zeta^2 x + 2\zeta^4 t)\hat{\sigma}_3} S(\zeta).
\end{align}

Evaluation of (\ref{seq}) and (\ref{Seq}) at $(x,t) = (0,0)$ and $(x,t) = (0,T)$ gives the expressions
\begin{equation}\label{Ssdef}   
   s(\zeta) = \mu_3(0,0, \zeta), \qquad S(\zeta) = \mu_1(0,0, \zeta) = \left(e^{2i\zeta^4T\hat{\sigma}_3}\mu_2(0,T,\zeta)\right)^{-1}.
\end{equation}
Hence, the functions $s(\zeta)$ and $S(\zeta)$ can be obtained from the evaluations at $x = 0$ respectively $t = T$ of the functions $\mu_3(x,0, \zeta)$ and $\mu_2(0,t,\zeta)$, which satisfy the linear integral equations
\begin{align}\label{mu3x0equation}
 & \mu_3(x,0,\zeta) = I + \int_{\infty}^{x} e^{i\zeta^2 (x' -x) \hat{\sigma}_3}(V_1\mu_3)(x',0,\zeta) dx',
	\\ \label{mu20tequation}
  &\mu_2(0,t,\zeta) = I + \int_{0}^{t} e^{2i\zeta^4 (t' -t) \hat{\sigma}_3}(V_2\mu_2)(0,t',\zeta) dt'.
\end{align}
By evaluating (\ref{V1explicit}) and (\ref{V2explicit}) at $t = 0$ and $x = 0$, respectively, we find that
\begin{equation}\label{V1initial}
V_1(x,0, \zeta) = \begin{pmatrix} -\frac{i}{2} |q_0|^2	&	\zeta q_0 e^{-i\int^x_{0} |q_0|^2 dx'} \\
\zeta \bar{q}_0 e^ {i \int^x_{0} |q_0|^2 dx'}	&	\frac{i}{2} |q_0|^2 \end{pmatrix}
\end{equation}
and
\begin{align}\label{V2boundary}
& V_2(0,t,\zeta) =
	\\ \nonumber
& \begin{pmatrix} -i\zeta^2 |g_0|^2 - \frac{3i}{4}|g_0|^4 - \frac{1}{2}(\bar{g}_1g_0 - \bar{g}_0g_1)		&	\left(2\zeta^3 g_0+ i\zeta g_1 + \zeta g_0|g_0|^2\right)e^ {-2i\int^{t}_{0} \Delta_2(0, t')dt'}	\\
 \left(2\zeta^3 \bar{g}_0 - i\zeta \bar{g}_1 + \zeta \bar{g}_0 |g_0|^2 \right)e^{2i\int^{t}_{0} \Delta_2(0, t')dt'}	&	i\zeta^2 |g_0|^2 + \frac{3i}{4} |g_0|^4 +  \frac{1}{2}(\bar{g}_1g_0 - \bar{g}_0g_1)	\end{pmatrix},
\end{align}
where $q_0(x) = q(x,0)$, $g_0(t) = q(0,t)$, and $g_1(t) = q_x(0,t)$ are the initial and boundary values of $q(x,t)$, and
$$\Delta_2(0,t) = \frac{3}{4} |g_0|^4 - \frac{i}{2}(\bar{g}_1g_0 - \bar{g}_0g_1).$$
These expressions for $V_1(x,0, \zeta)$ and $V_2(0,t,\zeta)$ contain only $q_0(x)$ and $\{g_0(t), g_1(t)\}$, respectively. Therefore, the integral equation (\ref{mu3x0equation}) determining $s(\zeta)$ is defined in terms of the initial data $q_0(x)$, and the integral equation (\ref{mu20tequation}) determining $S(\zeta)$ is defined in terms of the boundary values $g_0(t)$ and $g_1(t)$.

\subsection{Symmetries}
\begin{proposition}
For $j= 1,2,3$, the function $\mu(x,t,\zeta) = \mu_j(x,t,\zeta)$ satisfies the symmetry relations
\begin{align}\label{musymmetries1}
  \mu_{11}(x,t,\zeta) = \overline{\mu_{22}(x,t,\bar{\zeta})}, \qquad \mu_{21}(x,t,\zeta) = \overline{\mu_{12}(x,t, \bar{\zeta})},
\end{align}
as well as
\begin{align}\label{musymmetries2}
  \mu_{12}(x, t, -\zeta) = -\mu_{12}(x, t, \zeta), \qquad \mu_{11}(x, t, -\zeta) = \mu_{11}(x, t,\zeta),
  		\\\nonumber
    \mu_{21}(x, t, -\zeta) = -\mu_{21}(x, t, \zeta), \qquad \mu_{22}(x, t, -\zeta) = \mu_{22}(x, t, \zeta).
\end{align}
\end{proposition}
\proofbegin
To show (\ref{musymmetries1}) we introduce the following notation: for a $2\times 2$ matrix $A$, the $2\times 2$ matrix $TA$ is defined by
$$TA = \begin{pmatrix} 
      \bar{a}_{22} & \bar{a}_{21} \\
      \bar{a}_{12} & \bar{a}_{11} \\
   \end{pmatrix}
\quad \text{where} \quad
A = \begin{pmatrix} 
      a_{11} & a_{12} \\
      a_{21} & a_{22} \\
   \end{pmatrix}.$$
This operation has the property that $T(AB) = (TA)(TB)$ for any $2\times2$ matrices $A$ and $B$. In particular, $Te^A = e^{TA}$. 
Applying $T$ to equation (\ref{laxdiffform}) for $\mu$,
\begin{equation}\label{laxdiffform2} 
  d\left(e^{i(\zeta^2 x + 2\zeta^4 t)\hat{\sigma}_3} \mu(x,t,\zeta) \right) = e^{i(\zeta^2 x + 2\zeta^4 t)\hat{\sigma}_3} V(x,t,\zeta)\mu(x,t,\zeta),
\end{equation}  
we obtain
\begin{equation}\label{laxdiffformsquared} 
d\left(e^{i(\bar{\zeta}^2 x + 2\bar{\zeta}^4 t)\hat{\sigma}_3} (T\mu)(x,t,\zeta) \right) = e^{i(\bar{\zeta}^2 x + 2\bar{\zeta}^4 t)\hat{\sigma}_3} (TV)(x,t,\zeta)(T\mu)(x,t,\zeta).
\end{equation}
Since $TQ = Q$, definition (\ref{Udef}) of $U$ gives
$$(TU)(\zeta) = TU_1 dx + TU_2 dt =  \bar{\zeta} Q dx+ (-i\bar{\zeta}^2 Q^2\sigma_3 + 2\bar{\zeta}^3 Q - i\bar{\zeta} Q_x\sigma_3 + \bar{\zeta} Q^3)dt.$$
From this expression it is clear that $(TU)(\bar{\zeta}) = U(\zeta)$ and so
$$(TV)(\bar{\zeta}) = (e^{-i\int^{(x,t)}_{(0, 0)} \Delta \hat{\sigma}_3}(TU)( \bar{\zeta})) - i\Delta \sigma_3 = V(\zeta).$$
Hence, replacing $\zeta$ by $\bar{\zeta}$ in (\ref{laxdiffformsquared}), we infer that both $(T\mu)(x,t, \bar{\zeta})$ and $\mu(x,t,\zeta)$ satisfy equation (\ref{laxdiffform2}). If $\mu = \mu_j$, $j = 1,2,3$, then $\mu$ satisfies the initial condition $\mu(x_j, t_j, \zeta) = I$ for all $\zeta$ for which $\mu$ is defined. It follows that $(T\mu)(x_j,t_j, \bar{\zeta}) = TI = I$ satisfies the same initial condition. Thus, $(T\mu)(x,t, \bar{\zeta})$ and $\mu(x,t,\zeta)$ are equal, which in components is exactly (\ref{musymmetries1}).

To show (\ref{musymmetries2}) we define, for a $2\times 2$ matrix $A$, the $2\times 2$ matrix $PA$ by
\begin{equation*}\label{Pdef}
PA = \begin{pmatrix} 
      a_{11} & -a_{12} \\
      -a_{21} & a_{22} \\
   \end{pmatrix}
\quad \text{where} \quad
A = \begin{pmatrix} 
      a_{11} & a_{12} \\
      a_{21} & a_{22} \\
   \end{pmatrix}.
 \end{equation*}  
Just like $T$, this operation also has the properties that $P(AB) = (PA)(PB)$ for any $2\times2$ matrices $A$ and $B$, $P(\exp(A)) = \exp(PA)$, and $P(i\sigma_3) = i\sigma_3$. Moreover, since $PQ = -Q$, it is easily checked that $(PV)(-\zeta) = V(\zeta)$. The same kind of argument that led to (\ref{musymmetries1}) now gives (\ref{musymmetries2}).
\proofend

\subsection{The functions $s(\zeta)$ and $S(\zeta)$}\label{abABsubsec}
If $\psi(x,t,\zeta)$ satisfies (\ref{psilax}), it follows that $\det \psi$ is independent of $x$ and $t$. Hence, since $\det D(x,t) = 1$, the determinant of the function $\mu$ corresponding to $\psi$ according to (\ref{psimurelation}) is also independent of $x$ and $t$. In particular, for $\mu_j$, $j = 1,2,3$, evaluation of $\det \mu_j$ at $(x_j, t_j)$ shows that
$$\det \mu_j = 1, \quad j= 1,2,3.$$
In particular,
$$\det s(\zeta) = \det S(\zeta) = 1.$$

It follows from (\ref{musymmetries1}) that
$$s_{11}(\zeta) = \overline{s_{22}(\bar{\zeta})}, \quad s_{21}(\zeta) = \overline{s_{12}(\bar{\zeta})}, \quad S_{11}(\zeta) = \overline{S_{22}(\bar{\zeta})}, \quad S_{21}(\zeta) = \overline{S_{12}(\bar{\zeta})},$$
so that we may use the following notation for $s$ and $S$,
$$s(\zeta) = \begin{pmatrix} \overline{a(\bar{\zeta})} & b(\zeta) \\
 \overline{b(\bar{\zeta})} 	&	a(\zeta) \end{pmatrix}, \qquad 
S(\zeta) = \begin{pmatrix} \overline{A(\bar{\zeta})} & B(\zeta) \\
 \overline{B(\bar{\zeta})} 	&	A(\zeta) \end{pmatrix}.$$
The symmetries in (\ref{musymmetries2}) implies that $a(\zeta)$ and $A(\zeta)$ are even functions of $\zeta$, whereas $b(\zeta)$ and $B(\zeta)$ are odd functions of $\zeta$, that is,
\begin{equation}\label{abABevenoddsymmetries}  
  a(-\zeta) = a(\zeta), \quad b(-\zeta) = -b(\zeta), \quad A(-\zeta) = A(\zeta), \quad B(-\zeta) = -B(\zeta).
\end{equation}

The definitions of $\mu_j(0,t,\zeta)$, $j = 1,2$, and of $\mu_2(x,0,\zeta)$ imply that these functions have the larger domains of boundedness
$$\mu_1(0,t,\zeta) = \left(\mu_1^{(24)}(0,t,\zeta), \mu_1^{(13)}(0,t,\zeta)\right),$$
$$\mu_2(0,t,\zeta) = \left(\mu_2^{(13)}(0,t,\zeta), \mu_2^{(24)}(0,t,\zeta)\right),$$
$$\mu_2(x,0,\zeta) = \left(\mu_2^{(12)}(x,0,\zeta), \mu_2^{(34)}(x,0,\zeta)\right).$$
The definitions of $s(\zeta)$ and $S(\zeta)$ imply
$$\begin{pmatrix}b(\zeta) \\
a(\zeta) \end{pmatrix} = \mu_3^{(12)}(0,0,\zeta), \qquad \begin{pmatrix} -e^{-4i\zeta^4 T} B(\zeta) \\
\overline{A(\bar{\zeta})} \end{pmatrix} = \mu_2^{(24)}(0, T, \zeta).$$
Let us summarize the properties of the spectral functions.
\begin{itemize}
\item  $a(\zeta)$ and $b(\zeta)$ are defined for $\{\zeta \in \C | \text{\upshape Im}\, \zeta^2 \geq 0\}$ and analytic for $\{\zeta \in \C | \text{\upshape Im}\, \zeta^2 > 0\}$.
\item $a(\zeta)\overline{a(\bar{\zeta})} -  b(\zeta)\overline{b(\bar{\zeta})} = 1, \qquad \zeta^2 \in \R^2$.
\item $a(\zeta) = 1 + \left(\frac{1}{\zeta}\right), \qquad b(\zeta) = \left(\frac{1}{\zeta}\right), \qquad \zeta \to \infty, \quad \text{\upshape Im}\, \zeta^2 \geq 0.$

\item $A(\zeta)$ and $B(\zeta)$ are entire functions bounded for $\{\zeta \in \C | \text{\upshape Im}\, \zeta^4 \geq 0\}$. If $T = \infty$, the functions $A(\zeta)$ and $B(\zeta)$ are defined only for $\{\zeta \in \C | \text{\upshape Im}\, \zeta^4 \geq 0\}$.
\item $A(\zeta)\overline{A(\bar{\zeta})} -  B(\zeta)\overline{B(\bar{\zeta})} = 1, \qquad \zeta \in \C \quad (\zeta^4 \in \R \text{  if  } T = \infty)$.
\item  $A(\zeta) = 1 + O\left(\frac{1}{\zeta}\right),
%+ O\left(\frac{e^{4i\zeta^4T}}{\zeta}\right), 
\qquad B(\zeta) = O\left(\frac{1}{\zeta}\right),
%+ O\left(\frac{e^{4i\zeta^4T}}{\zeta}\right), 
\qquad \zeta \to \infty, \quad \text{\upshape Im}\, \zeta^4 \geq 0.$
\end{itemize}

All of these properties follow from the analyticity and boundedness of $\mu_3(x,0,\zeta)$ and $\mu_1(0,t,\zeta)$, from the conditions of unit determinant, and from the large $\zeta$ asymptotics of these eigenfunctions.

\subsection{The Riemann-Hilbert problem}
Equations (\ref{seq}) and (\ref{Seq}) can be rewritten in a form expressing the jump condition of a $2 \times 2$ RH problem. This involves only tedious but straightforward algebraic manipulations cf. \cite{F2002}. The final form is
$$M_-(x,t,\zeta) = M_+(x,t, \zeta)J(x,t,\zeta), \qquad \zeta^4 \in \R,$$
where the matrices $M_-$, $M_+$, and $J$ are defined by
\begin{align}\label{MplusMminusdef}
M_+ = \left(\frac{\mu_2^{(1)}}{a(\zeta)}, \mu_3^{(12)}\right), \quad \zeta \in D_1; \qquad
M_- = \left(\frac{\mu_1^{(2)}}{d(\zeta)}, \mu_3^{(12)}\right), \quad \zeta \in D_2;
		\\ \nonumber
M_+ = \left(\mu_3^{(34)}, \frac{\mu_1^{(3)}}{\overline{d(\bar{\zeta})}}\right), \quad \zeta \in D_3; \qquad
M_- = \left(\mu_3^{(34)}, \frac{\mu_2^{(4)}}{\overline{a(\bar{\zeta})}}\right), \quad \zeta \in D_4;
\end{align}
\begin{equation}\label{ddef}
  d(\zeta) =a(\zeta)\overline{A(\bar{\zeta})} -  b(\zeta)\overline{B(\bar{\zeta})}, \qquad \zeta \in \bar{D}_2;
\end{equation}
\begin{equation}
J(x,t,\zeta) = \left\{ \begin{array}{ll}
J_1 & \arg \zeta^2 = \frac{\pi}{2},  \\
J_2 = J_3 J_4^{-1} J_1 & \arg \zeta^2 = \pi, \\
J_3 & \arg \zeta^2 = \frac{3\pi}{2}, \\
J_4 & \arg \zeta^2 = 0, \\
\end{array} \right.
\end{equation}
\begin{figure}
\begin{center}
    \includegraphics[width=.3\textwidth]{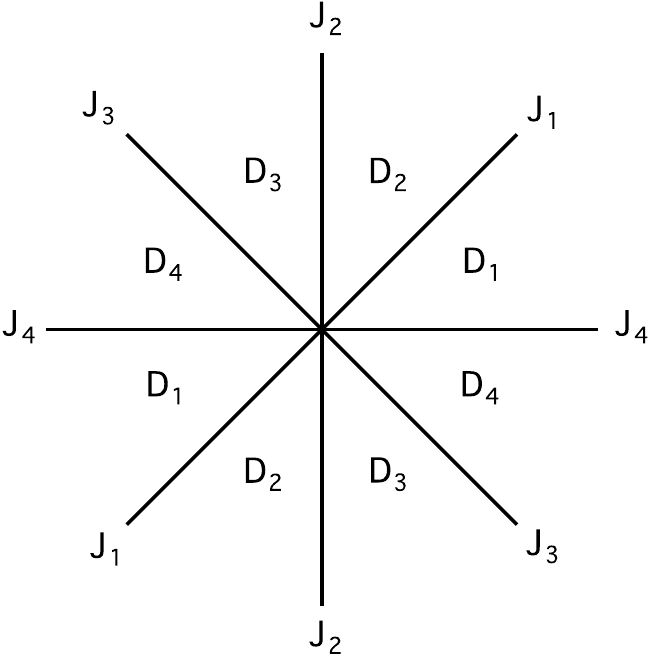} \\
     \begin{figuretext}\label{xtRHproblem.pdf}
        Illustration of the Riemann-Hilbert problem in the complex $\zeta$-plane.
     \end{figuretext}
     \end{center}
\end{figure}
with
\begin{equation}\label{J123def}
J_1 = \begin{pmatrix} 1	&	0 	\\
\Gamma(\zeta)e^{2i\theta(\zeta)}	&	1 \end{pmatrix}, \quad 
J_4 = \begin{pmatrix} 1	&	-\frac{b(\zeta)}{\overline{a(\bar{\zeta})}}e^{-2i\theta(\zeta)} 	\\
 \frac{\overline{b(\bar{\zeta})}}{a(\zeta)} e^{2i\theta(\zeta)}	&	\frac{1}{a(\zeta)\overline{a(\bar{\zeta})}} \end{pmatrix}, \quad 
J_3 = \begin{pmatrix} 1	&	- \overline{\Gamma(\bar{\zeta})}e^{-2i\theta(\zeta)} 	\\
0	&	1 \end{pmatrix};
\end{equation}
\begin{equation}\label{Gammadef}
\theta(\zeta) = \zeta^2 x + 2\zeta^4 t; \qquad \Gamma(\zeta) = \frac{ \overline{B(\bar{\zeta})}}{a(\zeta)d(\zeta)}, \qquad \zeta \in \bar{D}_2.
\end{equation}

The contour for this RH problem is depicted in Figure \ref{xtRHproblem.pdf}.

The matrix $M(x,t,\zeta)$ defined in (\ref{MplusMminusdef}) is in general a meromorphic function of $\zeta$ in $\C \setminus \{\zeta^4 \in \R\}$. The possible poles of $M$ are generated by the zeros of $a(\zeta)$, $d(\zeta)$, and by the complex conjugates of these zeros.

Since $a(\zeta)$ is an even function, each zero $\zeta_j$ of $a(\zeta)$ is accompanied by another zero at $-\zeta_j$. Similarly, each zero $\lambda_j$ of $d(\zeta)$ is accompanied by a zero at $-\lambda_j$. In particular, both $a(\zeta)$ and $d(\zeta)$ have an even number of zeros. 

\begin{assumption}\label{zerosassumption}\upshape We assume that
\item[(i)] $a(\zeta)$ has $2n$ simple zeros $\{\zeta_j\}_{j = 1}^{2n}$, $2n = 2n_1 + 2n_2$, such that $\zeta_j$, $j = 1, \dots, 2n_1$, lie in $D_1$ and $\zeta_j$, $j = 2n_1+1, \dots, 2n$, lie in $D_2$.

\item[(ii)] $d(\zeta)$ has $2\Lambda$ simple zeros $\{\lambda_j\}_1^{2\Lambda}$, such that $\lambda_j$, $j = 1, \dots, 2\Lambda$, lie in $D_2$.

\item[(iii)] None of the zeros of $a(\zeta)$ coincides with a zero of $d(\zeta)$.
\end{assumption}

In order to evaluate the associated residues we introduce the notation $[A]_1$ ($[A]_2$) for the first (second) column of a $2 \times 2$ matrix $A$ and we also write $\dot{a}(\zeta) = \frac{da}{d\zeta}$.
It holds that
\begin{align}\label{residue1}
\underset{\zeta_j}{\text{Res}} [M(x,t,\zeta)]_1 =& \frac{1}{\dot{a}(\zeta_j)b(\zeta_j)} e^{2i\theta(\zeta_j)} [M(x,t,\zeta_j)]_2, \qquad j = 1, \dots, 2n_1,
		\\\label{residue2}
\underset{\bar{\zeta}_j}{\text{Res}} [M(x,t,\zeta)]_2 =& \frac{1}{\overline{\dot{a}(\zeta_j)b(\zeta_j)}} e^{-2i\theta(\bar{\zeta}_j)} [M(x,t,\bar{\zeta}_j)]_1, \qquad j = 1, \dots, 2n_1,
	\\ \label{residue3}
\underset{\lambda_j}{\text{Res}} [M(x,t,\zeta)]_1 = &\frac{\overline{B(\bar{\lambda}_j)}}{a(\lambda_j)\dot{d}(\lambda_j)} e^{2i\theta(\lambda_j)} [M(x,t,\lambda_j)]_2, \qquad j = 1, \dots, 2\Lambda,
	\\\label{residue4}
\underset{\bar{\lambda}_j}{\text{Res}} [M(x,t,\zeta)]_2 =& \frac{B(\bar{\lambda}_j)}{\overline{a(\lambda_j)\dot{d}(\lambda_j)} }e^{-2i\theta(\bar{\lambda}_j)} [M(x,t,\bar{\lambda}_j)]_1, \qquad j = 1, \dots, 2\Lambda,
\end{align}
where $\theta(\zeta_j) = \zeta_j^2 x + 2\zeta_j^4 t.$

We shall prove (\ref{residue1}) and (\ref{residue3}); the proofs of (\ref{residue2}) and (\ref{residue4}) are analogous.

In order to derive equation (\ref{residue1}) we note that the second column of equation (\ref{seq}) is
$$\mu_3^{(12)} = a\mu_2^{(4)} + b\mu_2^{(1)}e^{-2i\theta}.$$
Recalling that $\mu_2$ is an entire function and evaluating this equation at $\zeta = \zeta_j$, $j = 1, \dots, 2n_1$, we find
$$\mu_3^{(12)}(\zeta_j) = b(\zeta_j)\mu_2^{(1)}(\zeta_j)e^{-2i\theta(\zeta_j)},$$
where, for simplicity of notation, we have suppressed the $x$ and $t$ dependence. Thus, since $[M]_1 = \mu_2^{(1)}/a$ in $D_1$, 
$$\underset{\zeta_j}{\text{Res}} [M]_1 = \frac{\mu_2^{(1)}(\zeta_j)}{\dot{a}(\zeta_j)} =  \frac{e^{2i\theta(\zeta_j)} \mu_3^{(12)}(\zeta_j)}{\dot{a}(\zeta_j)b(\zeta_j)},$$
which is equation (\ref{residue1}).

In order to derive equation (\ref{residue3}) we note that the first column of the equation $M_- = M_+J$ yields
$$a\mu_1^{(2)} = d\mu_2^{(1)} + \overline{B(\bar{\zeta})} e^{2i\theta} \mu_3^{(12)}.$$
Evaluating this equation at $\zeta = \lambda_j$ (each term has an analytic continuation for $\zeta \in D_2$) and using
$$\underset{\lambda_j}{\text{Res}} [M]_1 =
\frac{\mu_1^{(2)}(\lambda_j)}{\dot{d}(\lambda_j)}, \qquad [M(x, t, \lambda_j)]_2 = \mu_3^{(12)}(x,t,\lambda_j),$$
we find equation (\ref{residue3}).

\subsection{The inverse problem}\label{inverseproblemsubsec}
The inverse problem involves reconstructing the potential $q(x,t)$ from the spectral functions $\mu_j(x,t,\zeta)$, $j = 1,2,3$. We showed in section \ref{asymptoticsubsec} that $\Psi_1^{(o)} = \frac{i}{2}QD\sigma_3$ whenever 
$$\Psi = D + \frac{\Psi_1}{\zeta} + \frac{\Psi_2}{\zeta^2} + O\left(\frac{1}{\zeta^3}\right), \quad \zeta \to \infty,$$
is a solution of (\ref{Psilaxdiffform}). 
This implies that
\begin{equation}\label{recoverq}  
  q(x,t) = 2im(x,t)e^{2i\int^{(x,t)}_{(0, 0)} \Delta},
\end{equation}
where 
$$\mu = I + \frac{m^{(1)}}{\zeta} + \frac{m^{(2)}}{\zeta^2} + O\left(\frac{1}{\zeta^3}\right), \quad \zeta \to \infty,$$
is the corresponding solution of (\ref{laxdiffform}) related to $\Psi$ via (\ref{Psimurelation}), and we write $m(x,t)$ for $m^{(1)}_{12}(x,t)$.
From equation (\ref{recoverq}) and its complex conjugate, we obtain
$$rq = 4 |m|^2, \qquad r_xq - rq_x = 4\left(\bar{m}_xm - m_x\bar{m}\right) - 32i|m|^4.$$
Thus, we are able to express the one-form $\Delta$ defined in (\ref{Deltadef}) in terms of $m$ as
\begin{align}\label{Deltarecover}
  \Delta = 2 |m|^2 dx - \left(4|m|^4 + 2i\left(\bar{m}_xm -
m_x\bar{m}\right)\right)dt.
\end{align}
The inverse problem can now be solved as follows.
\begin{enumerate}
\item Use any of the three spectral functions $\mu_j$, $j =1,2,3$, to compute $m$ according to
$$m(x,t) = \lim_{\zeta \to \infty} \left(\zeta \mu_j(x,t, \zeta)\right)_{12}.$$
\item Determine $\Delta(x,t)$ from (\ref{Deltarecover}).
\item Finally, $q(x,t)$ is given by (\ref{recoverq}).
\end{enumerate}

\subsection{The global relation}
We now show that the spectral functions are not independent but satisfy an important global relation. Indeed, integrating the closed one-form $W = e^{i(\zeta^2 x + 2\zeta^4 t)\hat{\sigma}_3} V\mu$ in (\ref{laxdiffform}) with $\mu = \mu_3$ around the boundary of the domain $\{0 < x < \infty, 0 < t < T_0\}$, we get
\begin{align}\label{squareintegration}
&\int_{\infty}^0 e^{i\zeta^2 x' \hat{\sigma}_3} (V_1\mu_3)(x', 0, \zeta)dx' 
+  \int_0^{T_0}  e^{2i\zeta^4 t' \hat{\sigma}_3}  (V_2\mu_3)(0, t', \zeta)dt'
		\\ \nonumber
&+ e^{2i\zeta^4 T_0 \hat{\sigma}_3} \int_0^\infty   e^{i\zeta^2 x' \hat{\sigma}_3}  (V_1\mu_3)(x', T_0, \zeta)dx'
= \lim_{X \to \infty} e^{i\zeta^2 X \hat{\sigma}_3} \int_0^{T_0}  e^{2i\zeta^4 t' \hat{\sigma}_3}  (V_2\mu_3)(X, t', \zeta)dt'.
\end{align}

Using that $s(\zeta) = \mu_3(0,0, \zeta)$ it follows from (\ref{mu3x0equation}) that the first term of this equation equals $s(\zeta) - I$. Equation (\ref{seq}) evaluated at $x = 0$ gives
$$\mu_3(0,t',\zeta) = \mu_2(0,t',\zeta)e^{-2i\zeta^4 t' \hat{\sigma}_3} s(\zeta).$$
Thus,
$$e^{2i\zeta^4 t' \hat{\sigma}_3}  (V_2\mu_3)(0, t', \zeta) = \left[e^{2i\zeta^4 t' \hat{\sigma}_3}  (V_2\mu_2)(0, t', \zeta)\right]s(\zeta).$$
This equation, together with (\ref{mu20tequation}), implies that the second term of (\ref{squareintegration}) is
$$\int_0^{T_0} e^{2i\zeta^4 t' \hat{\sigma}_3}  (V_2\mu_3)(0, t', \zeta)dt'  = \left[e^{2i\zeta^4 T_0 \hat{\sigma}_3} \mu_2(0,T_0, \zeta) - I\right]s(\zeta).$$
Hence, assuming that $q$ has sufficient decay as $x \to \infty$, equation (\ref{squareintegration}) becomes
\begin{equation}\label{squareintegration2} 
 -I + S(T_0, \zeta)^{-1}s(\zeta) + e^{2i\zeta^4 T_0 \hat{\sigma}_3} \int_0^\infty  e^{i\zeta^2 x' \hat{\sigma}_3}  (V_1\mu_3)(x', T_0, \zeta)dx'  = 0,
\end{equation}
where the first and second columns of this equation are valid for $\zeta^2$ in the lower and the upper half-plane, respectively, and $S(T_0, \zeta)$ is defined by
$$S(T_0, \zeta) = \left(e^{2i\zeta^4T_0\hat{\sigma}_3}\mu_2(0,T_0,\zeta)\right)^{-1}.$$
Letting $T_0 = T$ and noting that $S(\zeta) = S(T, \zeta)$, equation (\ref{squareintegration2}) becomes
$$ -I + S(\zeta)^{-1}s(\zeta) + e^{2i\zeta^4 T \hat{\sigma}_3} \int_0^\infty  e^{i\zeta^2 x' \hat{\sigma}_3}  (V_1\mu_3)(x', T, \zeta)dx'  = 0.$$
The $(12)$ component of this equation is
\begin{equation}\label{globalrelation}
   B(\zeta)a(\zeta) - A(\zeta)b(\zeta) = e^{4i\zeta^4 T} c^+(\zeta), \qquad \text{\upshape Im}\, \zeta^2 \geq 0,
\end{equation}
where
$$c^+(\zeta) = \int_0^\infty e^{2i\zeta^2 x' }  (V_1\mu_3)_{12}(x', T, \zeta)dx' .$$
Equation (\ref{globalrelation}) is the global relation.

\section{The spectral functions}\label{spectralfunctionssec}\nequation
The analysis of section \ref{spectralanalysissec} motivates the following definitions for the spectral functions.

\begin{definition}[The spectral functions $a(\zeta)$ and $b(\zeta)$]\label{abdef}\upshape

Given $q_0(x) \in S(\R^+)$, we define the map
$$\mathbb{S}:\{q_0(x)\} \to \{a(\zeta), b(\zeta)\}$$
by
$$\begin{pmatrix} b(\zeta) \\
 a(\zeta) \end{pmatrix} = [\mu_3(0,\zeta)]_2, \qquad \text{\upshape Im}\, \zeta^2 \geq 0,$$
where $\mu_3(x,\zeta)$ is the unique solution of the Volterra linear integral equation 
$$\mu_3(x,\zeta) = I + \int_{\infty}^{x} e^{i\zeta^2 (x' -x) \hat{\sigma}_3}V_1(x',0,\zeta) \mu_3(x',\zeta) dx',$$
and $V_1(x,0,\zeta)$ is given in terms of $q_0(x)$ by equation (\ref{V1initial}).
\end{definition}

\begin{proposition}\label{abprop}
The spectral functions $a(\zeta)$ and $b(\zeta)$ have the properties
\begin{enumerate}
\item[(i)] $a(\zeta)$ and $b(\zeta)$ are analytic for $\text{\upshape Im}\, \zeta^2 > 0$ and continuous and bounded for $\text{\upshape Im}\, \zeta^2 \geq 0$.
\item[(ii)] $a(\zeta) = 1 + O(1/\zeta), \quad b(\zeta) = O(1/\zeta), \qquad \zeta \to \infty, \quad \text{\upshape Im}\, \zeta^2 \geq 0$.
\item[(iii)] $a(\zeta)\overline{a(\bar{\zeta})} - 
b(\zeta)\overline{b(\bar{\zeta})} = 1, \quad \zeta^2 \in \R.$
\item[(iv)] $a(-\zeta) = a(\zeta), \quad b(-\zeta) = -b(\zeta), \qquad \text{\upshape Im}\, \zeta^2 \geq 0$.
\item[(v)] The map $\Q:\{a(\zeta), b(\zeta)\} \mapsto \{q_0(x)\}$, inverse to $\mathbb{S}$, is defined by
\begin{equation}\label{recoverq0}
q_0(x) = 2im(x)e^{4i\int^x_0 |m(x')|^2 dx'}, \qquad m(x) = \lim_{\zeta \to \infty} \left(\zeta M^{(x)}(x, \zeta)\right)_{12},
\end{equation}
where $M^{(x)}(x,\zeta)$ is the unique solution of the following RH problem
\begin{itemize}
\item $M^{(x)}(x,\zeta) = \left\{ \begin{array}{ll}
M_-^{(x)}(x,\zeta) &  \text{\upshape Im}\, \zeta^2 \leq 0 \\
M_+^{(x)}(x,\zeta) &  \text{\upshape Im}\, \zeta^2 \geq 0 \\
\end{array} \right.$ 

is a sectionally meromorphic function.

\item $M_-^{(x)}(x,\zeta) = M_+^{(x)}(x,\zeta) J^{(x)}(x,\zeta), \qquad \zeta^2 \in \R,$

where
\begin{equation}\label{Jxdef}
J^{(x)}(x, \zeta) = \begin{pmatrix} 1 & -\frac{b(\zeta)}{\overline{a(\bar{\zeta})}}e^{-2i\zeta^2x} \\
\frac{ \overline{b(\bar{\zeta})}}{a(\zeta)}e^{2i\zeta^2x}	& 	\frac{1}{a(\zeta)\overline{a(\bar{\zeta})}}\end{pmatrix}, \qquad \zeta^2 \in \R.
\end{equation}
\item $M^{(x)}(x,\zeta) = I + O\left(\frac{1}{\zeta}\right), \qquad \zeta \to \infty.$
\item $a(\zeta)$ has $2n$ simple zeros $\{\zeta_j\}_{j = 1}^{2n}$, $2n = 2n_1 + 2n_2$, such that $\zeta_j$, $j = 1, \dots, 2n_1$, lie in $D_1$ and $\zeta_j$, $j = 2n_1+1, \dots, 2n$, lie in $D_2$
\item The first column of $M^{(x)}_+$ has simple poles at $\zeta = \zeta_j$, $j = 1, \dots, 2n$, and the second column of $M^{(x)}_-$ has simple poles at $\zeta = \bar{\zeta}_j$, $j = 1, \dots, 2n$.
The associated residues are given by
\begin{align}\label{xpartresidues}
\underset{\zeta_j}{\text{\upshape Res}} [M^{(x)}(x,\zeta)]_1 = \frac{e^{2i\zeta_j^2 x}}{\dot{a}(\zeta_j)b(\zeta_j)} [M^{(x)}(x, \zeta_j)]_2,  \qquad j = 1, \dots, 2n,
		\\ \label{xpartresiduesbar}
\underset{\bar{\zeta}_j}{\text{\upshape Res}} [M^{(x)}(x,\zeta)]_2 = \frac{ e^{-2i\bar{\zeta}_j^2 x}}{\overline{\dot{a}(\zeta_j)b(\zeta_j)}} [M^{(x)}(x, \bar{\zeta}_j)]_1,  \qquad j = 1, \dots, 2n.
\end{align}

\end{itemize}

\item[(vi)] We have
$$\mathbb{S}^{-1} = \Q.$$

\end{enumerate}

\end{proposition}
\proofbegin
$(i)-(iv)$ follow from the discussion in section \ref{abABsubsec}; the derivation of $(v)$ and $(vi)$ is given in appendix \ref{xinverseappendix}.
\proofend

\begin{definition}[The spectral functions $A(\zeta)$ and $B(\zeta)$]\label{ABdef}\upshape
Let $g_0(t)$ and $g_1(t)$ be smooth functions. The map
$$\tilde{\mathbb{S}}:\{g_0(t), g_1(t)\} \to \{A(\zeta), B(\zeta)\}$$
is defined by
$$\begin{pmatrix} B(\zeta) \\ A(\zeta) \end{pmatrix}
= [\mu_1(0, \zeta)]_2,$$
where $\mu_1(t, \zeta)$ is the unique solution of Volterra linear integral equation
$$\mu_1(t,\zeta) = I + \int_{T}^{t} e^{2i\zeta^4 (t' -t) \hat{\sigma}_3}V_2(0,t',\zeta) \mu_1(t',\zeta) dt',$$
and $V_2(0,t,\zeta)$ is given in terms of $\{g_0(t), g_1(t)\}$ by equation (\ref{V2boundary}).
\end{definition}

\begin{proposition}\label{ABprop}
The spectral functions $A(\zeta)$ and $B(\zeta)$ have the following properties.
\begin{enumerate}
\item[(i)] $A(\zeta)$ and $B(\zeta)$ are entire functions bounded for $\text{\upshape Im}\, \zeta^4 \geq 0$. If $T = \infty$, the functions $A(\zeta)$ and $B(\zeta)$ are defined only for $\text{\upshape Im}\, \zeta^4 \geq 0$.
\item[(ii)] $A(\zeta) = 1 + O(1/\zeta), \quad B(\zeta) = O(1/\zeta), \qquad \zeta \to \infty, \quad \text{\upshape Im}\, \zeta^4 \geq 0$.
\item[(iii)] $A(\zeta)\overline{A(\bar{\zeta})} - B(\zeta)\overline{B(\bar{\zeta})} = 1, \qquad \zeta \in \C$ ($\zeta^4 \in \R$ if $T = \infty$).
\item[(iv)] $A(-\zeta) = A(\zeta), \quad B(-\zeta) = -B(\zeta)$.
\item[(v)] The map $\tilde{\Q}:\{A(\zeta), B(\zeta)\} \mapsto \{g_0(t), g_1(t)\}$, inverse to $\tilde{\mathbb{S}}$, is defined by
\begin{align}\label{recoverg0g1}
& g_0(t) = 2im^{(1)}_{12}(t)e^{2i\int^{t}_0 \Delta_2(t')dt'},
		\\ \nonumber
  & g_1(t) = \left(4 m^{(3)}_{12}(t) + |g_0(t)|^2 m^{(1)}_{12}(t)\right)e^{2i\int^{t}_0 \Delta_2(t')dt'} + ig_0(t) \left(2m^{(2)}_{22}(t) +  |g_0(t)|^2\right),
\end{align}
where
$$\Delta_2(t) = 4\left|m^{(1)}_{12}\right|^4 + 8 \left(\text{\upshape Re}\,\left[m^{(1)}_{12} \bar{m}^{(3)}_{12}\right] - \left|m^{(1)}_{12}\right|^2\text{\upshape Re}\,\left[m^{(2)}_{22}\right]\right),$$
and the functions $m^{(1)}(t)$, $m^{(2)}(t)$, and $m^{(3)}(t)$ are determined by the asymptotic expansion
$$M^{(t)}(t, \zeta) = I + \frac{m^{(1)}(t)}{\zeta} + \frac{m^{(2)}(t)}{\zeta^2}+ \frac{m^{(3)}(t)}{\zeta^3} + O\left(\frac{1}{\zeta^4}\right), \quad \zeta \to \infty,$$
where $M^{(t)}(t,\zeta)$ is the unique solution of the following RH problem
\begin{itemize}
\item $M^{(t)}(t,\zeta) = \left\{ \begin{array}{ll}
M_-^{(t)}(t,\zeta) &  \text{\upshape Im}\, \zeta^4 \leq 0 \\
M_+^{(t)}(t,\zeta) &  \text{\upshape Im}\, \zeta^4 \geq 0 \\
\end{array} \right.$

is a sectionally meromorphic function.
\item $M_-^{(t)}(t,\zeta) = M_+^{(t)}(t,\zeta) J^{(t)}(t,\zeta), \qquad \zeta^4 \in \R,$

where
\begin{equation}\label{Jtdef}
J^{(t)}(t, \zeta) = \begin{pmatrix} 1 & -\frac{B(\zeta)}{\overline{A(\bar{\zeta})}}e^{-4i\zeta^4t} \\
\frac{ \overline{B(\bar{\zeta})}}{A(\zeta)}e^{4i\zeta^4 t}	& 	\frac{1}{A(\zeta)\overline{A(\bar{\zeta})}}\end{pmatrix}, \qquad \zeta^4 \in \R.
\end{equation}
\item $M^{(t)}(t,\zeta) = I + O\left(\frac{1}{\zeta}\right), \qquad \zeta \to \infty.$
\item We assume that $A(\zeta)$ has $2N$ simple zeros $\{z_j\}_{j = 1}^{2N}$ such that $\text{\upshape Im}\, z_j^4 > 0$, $j = 1,\dots,2N$. 
\item The first column of $M^{(t)}_+$ has simple poles at $\zeta = z_j$, $j = 1, \dots, 2N$, and the second column of $M^{(t)}_-$ has simple poles at $\zeta = \bar{z}_j$, $j = 1, \dots, 2N$.
The associated residues are given by
\begin{align}\label{tpartresidues}
& \underset{z_j}{\text{\upshape Res}}  [M^{(t)}(t,\zeta)]_1 = \frac{e^{4iz_j^4 t}}{\dot{A}(z_j)B(z_j)} [M^{(t)}(t, z_j)]_2, \qquad j = 1, \dots, 2N,
		\\ \label{tpartresiduesbar}
& \underset{\bar{z}_j}{\text{\upshape Res}} [M^{(t)}(x,\zeta)]_2 = \frac{ e^{-4i\bar{z}_j^4 t}}{\overline{\dot{A}(z_j)B(z_j)}} [M^{(t)}(t, \bar{z}_j)]_1,  \qquad j = 1, \dots, 2N.
\end{align}

\end{itemize}

\item[(vi)] We have
$$\tilde{\mathbb{S}}^{-1} = \tilde{\Q}.$$

\end{enumerate}

\end{proposition}
\proofbegin
$(i)-(iv)$ follow from the discussion in section \ref{abABsubsec}; the derivation of $(v)$ and $(vi)$ is given in appendix \ref{tinverseappendix}.
\proofend

\section{The Riemann-Hilbert problem}\label{RHsec}\nequation

\begin{theorem}\label{RHtheorem}
Let $q_0(x) \in S(\R^+)$. Suppose that the functions $g_0(t)$ and $g_1(t)$ are compatible with $q_0(x)$ at $x=t=0$. Define the spectral functions $a(\zeta)$, $b(\zeta)$, $A(\zeta)$, and $B(\zeta)$ in terms of $q_0(x)$, $g_0(t)$, and $g_1(t)$ according to Definitions \ref{abdef} and \ref{ABdef}. Suppose that the global relation (\ref{globalrelation}) is satisfied for some $c^+(\zeta)$ which is analytic for $\text{\upshape Im}\, \zeta^2 > 0$, continuous and bounded for $\text{\upshape Im}\, \zeta^2 \geq 0$, and such that $c^+(\zeta) = O(1/\zeta)$, $\zeta \to \infty$; if $T= \infty$ the global relation is replaced by
$$B(\zeta)a(\zeta) - A(\zeta)b(\zeta) = 0, \qquad \zeta \in \bar{D}_1.$$
Assume that the possible zeros $\{\zeta_j\}_{1}^{2n}$ of $a(\zeta)$ and $\{\lambda_j\}_{1}^{2\Lambda}$ of $d(\zeta)$ are as in Assumption \ref{zerosassumption}. Define $M(x,t,\zeta)$ as the solution of the following $2\times 2$ matrix RH problem:
\begin{itemize}
\item $M$ is sectionally meromorphic in $\zeta \in \C \setminus \{\zeta^4 \in \R\}$.

\item The first column of $M$ has simple poles at $\zeta = \zeta_j$, $j = 1, \dots, 2n_1$, and at $\zeta = \lambda_j$, $j = 1, \dots, 2\Lambda$. The second column of $M$ has simple poles at $\zeta = \bar{\zeta}_j$, $j = 1, \dots, 2n_1$, and at $\zeta = \bar{\lambda}_j$, $j = 1, \dots, 2\Lambda$. The associated residues satisfy the relations in (\ref{residue1})-(\ref{residue4}).

\item $M$ satisfies the jump condition
$$M_-(x,t,\zeta) = M_+(x,t,\zeta)J(x,t,\zeta), \qquad \zeta^4 \in \R,$$
where $M$ is $M_-$ for $\text{\upshape Im}\, \zeta^4 \leq 0$, $M$ is $M_+$ for $\text{\upshape Im}\, \zeta^4 \geq 0$, and $J$ is defined in terms of $a,b,A$, and $B$ by equations (\ref{ddef})-(\ref{Gammadef}), see Figure \ref{xtRHproblem.pdf}.

\item $M(x,t,\zeta) = I + O\left(\frac{1}{\zeta}\right), \qquad \zeta \to \infty.$
\end{itemize}
Then $M(x,t,\zeta)$ exists and is unique.

Define $q(x,t)$ in terms of $M(x,t,\zeta)$ by
\begin{equation}\label{recoverqxt}
  q(x,t) = 2im(x,t)e^{2i\int^{(x,t)}_{(0, 0)} \Delta}, \qquad m(x,t) = \lim_{\zeta \to \infty} \left(\zeta M(x,t, \zeta)\right)_{12},
\end{equation}
$$\Delta = 2 |m|^2 dx - \left(4|m|^4 + 2i\left(\bar{m}_xm -
m_x\bar{m}\right)\right)dt.$$

Then $q(x,t)$ solves the DNLS equation (\ref{DNLS}). Furthermore,
$$q(x,0) = q_0(x), \qquad q(0,t) = g_0(t), \qquad \text{and} \qquad q_x(0,t) = g_1(t).$$
\end{theorem}
\proofbegin
In the case when $a(\zeta)$ and $d(\zeta)$ have no zeros, the unique solvability is a consequence of the following vanishing lemma.

\begin{lemma}[Vanishing lemma]\label{vanlemma}
The Riemann-Hilbert problem in Theorem \ref{RHtheorem} with the vanishing boundary condition
$$M(x, t, \zeta) \to 0 \quad \text{as} \quad \zeta \to \infty,$$
has only the zero solution.
\end{lemma}
\proofbegin
Assume that $M(x,t,\zeta)$ is a solution of the RH problem in Theorem \ref{RHtheorem} such that $M_\pm(x,t,\zeta) \to 0$ as $\zeta \to \infty$.
Let $A^\dagger$ denote the complex conjugate transpose of a matrix $A$ and define
\begin{align*}
& H_+(\zeta) = M_+ (\zeta)M_-^{\dagger}(-\bar{\zeta}), \qquad \text{\upshape Im}\, \zeta^4 \geq 0, 
	\\
& H_-(\zeta) = M_-(\zeta)M_+^{\dagger}(-\bar{\zeta}),  \qquad \text{\upshape Im}\, \zeta^4 \leq 0,
\end{align*}
where we suppress the $x$ and $t$ dependence for clarity.
$H_+(\zeta)$ and $H_-(\zeta)$ are analytic in $\{\zeta \in \C | \text{\upshape Im}\, \zeta^4 > 0\}$ and $\{\zeta \in \C | \text{\upshape Im}\, \zeta^4 < 0\}$, respectively. By the symmetry relations (\ref{abABevenoddsymmetries}), we infer that
$$J_3^\dagger(- \bar{\zeta}) = J_1(\zeta), \qquad J_4^\dagger(- \bar{\zeta}) = J_4(\zeta), \qquad J_2^\dagger(- \bar{\zeta}) = J_2(\zeta).$$
Since
$$H_+(\zeta) = M_+ (\zeta)J^{\dagger}(-\bar{\zeta})M_+^{\dagger}(-\bar{\zeta}), \qquad H_-(\zeta) = M_+(\zeta)J(\zeta)M_+^{\dagger}(-\bar{\zeta}), \qquad \zeta^4 \in \R,$$
this shows that $H_+(\zeta) = H_-(\zeta)$ for $\zeta^4 \in \R$. Therefore, $H_+(\zeta)$ and $H_-(\zeta)$ define an entire function vanishing at infinity, and so $H_+(\zeta)$ and $H_-(\zeta)$ are identically zero.
Now $J_2(is)$ is a hermitian matrix with unit determinant and $(22)$ entry $1$ for any $s \in \R$. Hence, $J_2(is)$, $s \in \R$, is a positive definite matrix. Since $H_-(s)$ vanishes identically for $s \in i\R$, i.e.
$$M_+(is)J_2(is)M_+^{\dagger}(is) = 0, \qquad s \in \R,$$
we deduce that $M_+(is) = 0$ for $s \in \R$. It follows that $M_+$ and $M_-$ vanish identically.

\proofend

If $a(\zeta)$ and $d(\zeta)$ have zeros this singular RH problem can be mapped to a regular one coupled with a system of algebraic equations \cite{F-I}.
Moreover, it follows from standard arguments using the dressing method \cite{Z-S1, Z-S2} that if $M$ solves the above RH problem and $q(x,t)$ is defined by (\ref{recoverqxt}), then $q(x,t)$ solves the DNLS equation (\ref{DNLS}).

\bigskip
{\it Proof that $q(x,0) = q_0(x)$.}
Define $M^{(x)}(x, \zeta)$ by
\begin{align}\label{MxM}
M^{(x)} &= M(x,0,\zeta), \qquad \zeta \in D_1 \cup D_4,
	\nonumber \\
M^{(x)} &= M(x,0,\zeta)J_1^{-1}(x,0,\zeta), \qquad \zeta \in D_2,
	\\ \nonumber
M^{(x)} &= M(x,0,\zeta)J_3(x,0,\zeta), \qquad \zeta \in D_3.
\end{align}
We first discuss the case where the sets $\{\zeta_j\}$ and $\{\lambda_j\}$ are empty. Then the function $M^{(x)}$ is analytic in $\C \setminus (\R \cup i\R)$. Furthermore,
$$M_-^{(x)}(x, \zeta) = M_+^{(x)}(x, \zeta)J^{(x)}(x, \zeta), \qquad \zeta^2 \in \R,$$
$$M^{(x)}(x, \zeta) = I + \left(\frac{1}{\zeta}\right), \qquad \zeta \to \infty,$$
where $J^{(x)}(x,\zeta)$ is defined in (\ref{Jxdef}). Thus according to (\ref{recoverq0}),
$$q_0(x) = 2im(x)e^{4i\int^x_0 |m(x')|^2 dx'}, \qquad m(x) = \lim_{\zeta \to \infty} \left(\zeta M^{(x)}(x, \zeta)\right)_{12},$$
Comparing this with equation (\ref{recoverqxt}) evaluated at $t = 0$, we conclude that $q_0(x) = q(x,0)$.

We now discuss the case when the sets $\{\zeta_j\}$ and $\{\lambda_j\}$ are not empty. The first column of $M(x,t,\zeta)$ has poles at $\{\zeta_j\}_{1}^{2n_1}$ for $\zeta \in D_1$ and has poles at $\{\lambda_j\}_1^{2\Lambda}$ for $\zeta \in D_2$. On the other hand, the first column of $M^{(x)}(x, \zeta)$ should have poles at $\{\zeta_j\}_1^{2n}$. We will now show that the transformation defined by (\ref{MxM}) map the former poles to the latter ones. Since $M^{(x)} = M(x,0,\zeta)$ for $\zeta \in D_1$, $M^{(x)}$ has poles at $\{\zeta_j\}_1^{2n_1}$ with the correct residue condition. Letting $M= (M_1, M_2)$, equation (\ref{MxM}) can be written as
\begin{equation}\label{MxM1Gamma}
M^{(x)}(x,\zeta) = (M_1(x,0,\zeta) - \Gamma(\zeta) e^{2i\zeta^2 x}M_2(x,0,\zeta), M_2(x,0,\zeta)),  \qquad \zeta \in D_2.
\end{equation}
The residue condition (\ref{residue3}) at $\lambda_j$ implies that $M^{(x)}$ has no poles at $\lambda_j$; on the other hand, equation (\ref{MxM1Gamma}) shows that $M^{(x)}$ has poles at $\{\zeta_j\}_{2n_1 + 1}^{2n}$ with residues given by
$$\underset{\zeta_j}{\text{Res}} [M^{(x)}(x,\zeta)]_1 = -\underset{\zeta_j}{\text{Res}} \Gamma(\zeta) e^{2i\zeta_j^2 x} [M^{(x)}(x,\zeta_j)]_2, \qquad j = 2n_1 +1, \dots, 2n.$$
Using the definition of $\Gamma(\zeta)$ and the equation $d(\zeta_j) = - b(\zeta_j)\overline{B(\bar{\zeta}_j)}$, this becomes the residue condition of (\ref{xpartresidues}). Similar considerations apply to $\bar{\zeta}_j$ and $\bar{\lambda}_j$.

\bigskip
{\it Proof that $q(0,t) = g_0(t)$ and $q_x(0,t) = g_1(t)$.}
Let $M^{(j)}(x,t,\zeta)$ denote $M(x,t,\zeta)$ for $\zeta \in D_j$, $j=1,\dots,4.$ Recall that $M$ satisfies
\begin{align}\label{M2M1J1etc}
M^{(2)} = M^{(1)}J_1, \qquad M^{(2)} = M^{(3)}J_2,
	\\	\nonumber
M^{(4)} = M^{(1)}J_4, \qquad M^{(4)} = M^{(3)}J_3,
\end{align}
on the respective parts of the boundary separating the $D_j$'s. Let $M^{(t)}(t,\zeta)$ be defined by
\begin{equation}\label{MtMG} 
  M^{(t)}(t,\zeta) = M(0,t,\zeta)G(t, \zeta),
\end{equation}
where $G$ is given by $G^{(j)}$ for $\zeta \in D_j$, $j=1,\dots,4.$ Suppose we can find matrices $G^{(1)}$ and $G^{(2)}$ holomorphic for $\text{\upshape Im}\, \zeta^2 > 0$ (and continuous for $\text{\upshape Im}\, \zeta^2 \geq 0$), matrices $G^{(3)}$ and $G^{(4)}$ holomorphic for $\text{\upshape Im}\, \zeta^2 < 0$ (and continuous for $\text{\upshape Im}\, \zeta^2 \leq 0$), which tend to $I$ as $\zeta \to \infty$, and which satisfy
\begin{align}
& J_1(0,t,\zeta) G^{(2)}(t,\zeta) = G^{(1)}(t,\zeta)J^{(t)}(t,\zeta), \qquad \zeta^2 \in i\R^+,
	\nonumber \\ \label{G1234def}
& J_3(0,t,\zeta) G^{(4)}(t,\zeta) = G^{(3)}(t,\zeta)J^{(t)}(t,\zeta), \qquad \zeta^2 \in i\R^-,
		\\ \nonumber
& J_4(0,t,\zeta) G^{(4)}(t,\zeta) = G^{(1)}(t,\zeta)J^{(t)}(t,\zeta), \qquad \zeta^2 \in \R^+,
\end{align}
where $J^{(t)}$ is the jump matrix defined in (\ref{Jtdef}). Then, since $J_2 = J_3 J_4^{-1} J_1$, it follows that
$$J_2(0,t,\zeta) G^{(2)}(t,\zeta) = G^{(3)}(t,\zeta)J^{(t)}(t,\zeta), \qquad \zeta^2 \in \R^-,$$
and equations (\ref{M2M1J1etc}) and (\ref{MtMG}) imply that $M^{(t)}$ satisfies the $RH$ problem defined in Proposition \ref{ABprop}. If the sets $\{\zeta_j\}$ and $\{\lambda_j\}$ are empty, this immediately yields the desired result.
We claim that such $G^{(j)}$ matrices are
$$G^{(1)} = \begin{pmatrix} \frac{a(\zeta)}{A(\zeta)}	&	c^+(\zeta)e^{4i\zeta^4 (T-t)}	\\
0	&	\frac{A(\zeta)}{a(\zeta)} \end{pmatrix}, \qquad
G^{(4)} = \begin{pmatrix} \frac{\overline{A(\bar{\zeta})}}{\overline{a(\bar{\zeta})}}	&	0	\\
 \overline{c^+(\bar{\zeta})}e^{-4i\zeta^4 (T-t)}	&	\frac{\overline{a(\bar{\zeta})}}{\overline{A(\bar{\zeta})}} \end{pmatrix},$$
$$G^{(2)} = \begin{pmatrix} d(\zeta)	&	\frac{-b(\zeta)e^{-4i\zeta^4 t}}{\overline{A(\bar{\zeta})}}	\\
0	&	\frac{1}{d(\zeta)} \end{pmatrix}, \qquad
G^{(3)} = \begin{pmatrix} \frac{1}{\overline{d(\bar{\zeta})}}		&	0	\\
\frac{ -  \overline{b(\bar{\zeta})}e^{4i\zeta^4t}}{A(\zeta)}	&	\overline{d(\bar{\zeta})} \end{pmatrix}.$$
We omit the verification that these $G^{(j)}$ matrices fulfill the requirements (\ref{G1234def}) as well as the verification of the residue conditions in the case of non-empty sets $\{\zeta_j\}$ and $\{\lambda_j\}$, since analogous arguments can be found in the proof of Theorem 4.1 in \cite{F-I-S}. 
\proofend

\appendix

\section{The $x$-inverse problem}\label{xinverseappendix}
\renewcommand{\theequation}{A.\arabic{equation}}\nequation
This appendix contains the derivation of statements $(v)$ and $(vi)$ of Proposition \ref{abprop}. 

\subsection{Formulation of RH problem}
Define solutions $\mu_3(x, \zeta)$ and $\mu_2(x, \zeta)$ of the $x$-part of equation (\ref{mulax}) by the Volterra linear integral equations
$$\mu_3(x,\zeta) = I + \int_{\infty}^{x} e^{i\zeta^2 (x' -x) \hat{\sigma}_3}V_1(x',0, \zeta) \mu_3(x',\zeta) dx',$$
$$\mu_2(x,\zeta) = I + \int_{0}^{x} e^{i\zeta^2 (x' -x)\hat{\sigma}_3}V_1(x',0, \zeta) \mu_2(x',\zeta) dx',$$
where $V_1(x,0, \zeta)$ is given by equation (\ref{V1initial}).
From (\ref{seq}) evaluated at $t =0$ it follows that 
\begin{align}\label{mu3mu2relation}
\mu_3(x,\zeta) &= \mu_2(x,\zeta)e^{-i\zeta^2 x\hat{\sigma}_3} s(\zeta)
= \mu_2(x,\zeta) \begin{pmatrix} \overline{a(\bar{\zeta})} & b(\zeta)e^{-2i\zeta^2 x} \\
 \overline{b(\bar{\zeta})}e^{2i\zeta^2 x} 	&	a(\zeta) \end{pmatrix}, \qquad \zeta^2 \in \R.
\end{align}
Let
$$M_-^{(x)} = \left([\mu_3]_1, \frac{[\mu_2]_2}{\overline{a(\bar{\zeta})}}\right), \qquad \text{\upshape Im}\, \zeta^2 \leq 0,$$
$$M_+^{(x)} = \left(\frac{[\mu_2]_1}{a(\zeta)}, [\mu_3]_2\right), \qquad \text{\upshape Im}\, \zeta^2 \geq 0.$$
Equation (\ref{mu3mu2relation}) can be rewritten as
$$M_-^{(x)}(x, \zeta) = M_+^{(x)}(x, \zeta)J^{(x)}(x, \zeta), \qquad \zeta^2 \in \R,$$
where $J^{(x)}$ is the jump matrix in (\ref{Jxdef}). If we can show that $M^{(x)}$ also fulfills the residue conditions (\ref{xpartresidues}) and (\ref{xpartresiduesbar}) at the possible simple zeros $\{\zeta_j\}_1^{2n}$ of $a(\zeta)$, it follows that $M^{(x)}$ satisfies the RH problem of Proposition \ref{abprop}.
To this end we notice that the second column of (\ref{mu3mu2relation}) is
$$[\mu_3]_2 = [\mu_2]_1b(\zeta)e^{-2i\zeta^2 x} + [\mu_2]_2 a(\zeta), \qquad \zeta^2 \in \R.$$
Since $\mu_2$ is an entire function of $\zeta$, we may evaluate this equation at $\zeta = \zeta_j$. This yields the relation
$$[\mu_3(x, \zeta_j)]_2 = [\mu_2(x, \zeta_j)]_1 b(\zeta_j)e^{-2i\zeta_j^2 x},$$
from which the first residue condition (\ref{xpartresidues}) follows. Condition (\ref{xpartresiduesbar}) follows similarly. To verify formula (\ref{recoverq0}) for the recovery of the potential $q_0(x)$ from $M^{(x)}$ one just has to apply the discussion of section \ref{inverseproblemsubsec} with $t$ set to zero.

\subsection{Unique solvability of RH problem}
Our next step is to show that the RH problem of Proposition \ref{abprop} $(v)$ admits a unique solution. In the case when $a(\zeta)$ has no zeros, the unique solvability is a consequence of the following vanishing lemma, whose proof is similar to that of Lemma \ref{vanlemma}.

\begin{lemma}[$x$-part vanishing lemma]
The Riemann-Hilbert problem in Proposition \ref{abprop} with the vanishing boundary conditions 
$$M^{(x)}(x, \zeta) \to 0 \quad \text{as} \quad \zeta \to \infty,$$
has only the zero solution. \proofend
\end{lemma}

In the case when $a(\zeta)$ has zeros, we transform the singular RH problem to a regular one as in \cite{F-I}. This shows that the RH problem of Proposition \ref{abprop} has a unique solution.

\subsection{Inverse of spectral map}
It remains to show that the map 
$$\Q:\{a(\zeta), b(\zeta)\} \mapsto \{q_0(x)\},$$ 
defined in Proposition \ref{abprop} $(v)$ is indeed the inverse of the spectral map $\mathbb{S}$.
In more detail, this problem is formulated as follows. Given $\{a(\zeta), b(\zeta)\}$, construct the jump matrix $J^{(x)}(x, \zeta)$ according to equation (\ref{Jxdef}) and let $M^{(x)}(x, \zeta)$ be the unique solution of the RH problem of Proposition \ref{abprop} $(v)$. Let $q_0(x)$ be the function defined by (\ref{recoverq0}), i.e.
$$q_0(x) = 2im(x)e^{4i\int^x_0 |m(x')|^2 dx'}, \qquad m(x) = \lim_{\zeta \to \infty} \left(\zeta M^{(x)}(x, \zeta)\right)_{12}.$$
Denote the spectral data corresponding to $q_0(x)$ by $\{a_0(\zeta), b_0(\zeta)\}$. We have to show that 
\begin{equation}\label{a0ab0bequal}  
  a_0(\zeta) = a(\zeta) \qquad \text{and} \qquad b_0(\zeta) = b(\zeta).
\end{equation}

Using the arguments of the dressing method \cite{Z-S1, Z-S2}, it is straightforward to prove that $M^{(x)}(x, \zeta)$ satisfies the $x$-part of (\ref{mulax}) with the potential $q_0(x)$ defined by (\ref{recoverq0}).
Equation (\ref{a0ab0bequal}) follows as in the case of the NLS equation (see appendix A of \cite{F-I-S}).

\section{The $t$-inverse problem}\label{tinverseappendix}
\renewcommand{\theequation}{B.\arabic{equation}}\nequation
This appendix contains the derivation of statements $(v)$ and $(vi)$ of Proposition \ref{ABprop}.

\subsection{Formulation of RH problem}
Define solutions $\mu_1(t, \zeta)$ and $\mu_2(t, \zeta)$ of the $t$-part of equation (\ref{mulax}) by the Volterra linear integral equations
$$\mu_1(t,\zeta) = I + \int_{T}^{t} e^{2i\zeta^4 (t' -t) \hat{\sigma}_3}V_2(0, t',\zeta)\mu_1(t',\zeta) dt',$$
$$\mu_2(t,\zeta) = I + \int_{0}^{t} e^{2i\zeta^4 (t' -t) \hat{\sigma}_3}V_2(0, t',\zeta)\mu_2(t',\zeta) dt',$$
where $V_2(0,t,\zeta)$ is given by equation (\ref{V2boundary}).
From (\ref{Seq}) evaluated at $x =0$ it follows that 
\begin{align}\label{mu1mu2relation}
  \mu_1(t,\zeta) = \mu_2(t,\zeta)  e^{-2i\zeta^4 t\hat{\sigma}_3} S(\zeta)
= \mu_2(t,\zeta) \begin{pmatrix} \overline{A(\bar{\zeta})} & B(\zeta)e^{-4i\zeta^4 t} \\
 \overline{B(\bar{\zeta})}e^{4i\zeta^4 t} 	&	A(\zeta) \end{pmatrix}, \qquad \zeta^4 \in \R.
\end{align}
Let
$$M_-^{(t)} = \left([\mu_1]_1, \frac{[\mu_2]_2}{\overline{A(\bar{\zeta})}} \right), \qquad \text{\upshape Im}\, \zeta^4 \leq 0,$$
$$M_+^{(t)} = \left(\frac{[\mu_2]_1}{A(\zeta)}, [\mu_1]_2\right), \qquad \text{\upshape Im}\, \zeta^4 \geq 0.$$
Equation (\ref{mu1mu2relation}) can be rewritten as
$$M_-^{(t)}(t, \zeta) = M_+^{(t)}(t, \zeta)J^{(t)}(t, \zeta), \qquad \zeta^4 \in \R,$$
where $J^{(t)}(t, \zeta)$ is the jump matrix defined in (\ref{Jtdef}).
To check that $J^{(t)}(t, \zeta)$ satisfies the residue conditions (\ref{tpartresidues}) and (\ref{tpartresiduesbar}) one proceeds as in the case of $J^{(x)}(x, \zeta)$.

\subsection{Unique solvability of RH problem}
Again the unique solvability of the RH problem of Proposition \ref{ABprop} is a consequence of a vanishing lemma in the case when $A(\zeta)$ has no zeros, while the case when $A(\zeta)$ has zeros can be transformed to a regular RH problem \cite{F-I}.

\begin{lemma}[$t$-part vanishing lemma]
The Riemann-Hilbert problem in Proposition \ref{ABprop} with the vanishing boundary conditions 
$$M^{(t)}(t, \zeta) \to 0 \quad \text{as} \quad \zeta \to \infty,$$
has only the zero solution. \proofend
\end{lemma}

\subsection{Recovering $g_0(t)$ and $g_1(t)$}
Our next step will be to establish formula (\ref{recoverg0g1}) for the recovery of $g_0(t)$ and $g_1(t)$ from $M^{(t)}$. Let $\mu(x,t,\zeta)$ be a solution of equation (\ref{laxdiffform}). From (\ref{orderonetpart}) we have
\begin{equation}\label{iQxsigma3etc}
iQ_x\sigma_3D = 4i\Psi_3^{(o)}\sigma_3 - iQ^2 \sigma_3\Psi_1^{(o)} + 2Q \Psi_2^{(d)} + Q^3 D,
\end{equation}
where
$$\Psi = D + \frac{\Psi_1}{\zeta} + \frac{\Psi_2}{\zeta^2} + \frac{\Psi_3}{\zeta^3} + O\left(\frac{1}{\zeta^4}\right), \quad \zeta \to \infty,$$
is the solution of (\ref{Psilaxdiffform}) related to $\mu$ via (\ref{Psimurelation}), i.e. 
$$\Psi
= \begin{pmatrix} d_1\mu_{11} & d_2 e^{2i\int^{(x,t)}_{(0, 0)} \Delta} \mu_{12} \\
d_1 e^{-2i\int^{(x,t)}_{(0, 0)} \Delta} \mu_{21}	&	d_2\mu_{22} \end{pmatrix},$$
where we have written 
$$D = \text{diag}(d_1, d_2), \qquad \mu = \begin{pmatrix} \mu_{11} & \mu_{12} \\
\mu_{21}	&	\mu_{22} \end{pmatrix}.$$ 
If we let
$$\mu = I + \frac{m^{(1)}}{\zeta} + \frac{m^{(2)}}{\zeta^2} + \frac{m^{(3)}}{\zeta^3} + O\left(\frac{1}{\zeta^4}\right), \qquad \zeta \to \infty,$$
then the $(12)$ entry of equation (\ref{iQxsigma3etc}) gives
%$$-iq_xd_2 = -4i(\Psi_3)_{12} - irq(\Psi_1)_{12} + 2q(\Psi_2)_{22} + rq^2 d_2.$$
%$$-iq_x = -4i e^{2i\int^{(x,t)}_{(0, 0)} \Delta} m^{(3)}_{12} - irqe^{2i\int^{(x,t)}_{(0, 0)} \Delta} m^{(1)}_{12} + 2q m^{(2)}_{22} + rq^2.$$
%Hence
\begin{equation}\label{qxfromm}  
  q_x = (4 m^{(3)}_{12} + rqm^{(1)}_{12})e^{2i\int^{(x,t)}_{(0, 0)} \Delta} + 2iq m^{(2)}_{22} + irq^2.
\end{equation}
Taking the complex conjugate one finds
\begin{equation}\label{rxfromm}  
  r_x = (4 \bar{m}^{(3)}_{12} + rq\bar{m}^{(1)}_{12})e^{-2i\int^{(x,t)}_{(0, 0)} \Delta} - 2ir \bar{m}^{(2)}_{22} - ir^2q.
\end{equation}
On the other hand, by (\ref{recoverq}) we have
\begin{equation}\label{qrfromm}
q(x,t) = 2im^{(1)}_{12}e^{2i\int^{(x,t)}_{(0, 0)} \Delta}, \qquad r(x,t) = -2i \bar{m}^{(1)}_{12}e^{-2i\int^{(x,t)}_{(0, 0)} \Delta}.
\end{equation}
From (\ref{qxfromm})-(\ref{qrfromm}) it follows that
\begin{equation}\label{rxqrqxfromm}
r_x q -r q_x = 4i \text{Re}\,[m^{(1)}_{12}(4 \bar{m}^{(3)}_{12} + rq\bar{m}^{(1)}_{12})] - 4irq \text{Re}\, [m^{(2)}_{22}] - 2ir^2q^2.
\end{equation}
This shows that the coefficient $\Delta_2 = \frac{3}{4} r^2q^2 - \frac{i}{2}(r_xq - rq_x)$ of $dt$ in the differential form $\Delta$ as defined in (\ref{Deltadef}) can be expressed as
\begin{equation}\label{Delta2expression}
 \Delta_2 = -\frac{1}{4} r^2q^2 + 2 \text{Re}\,\left[m^{(1)}_{12}\left(4 \bar{m}^{(3)}_{12} + rq\bar{m}^{(1)}_{12}\right)\right] - 2rq \text{Re}\, \left[m^{(2)}_{22}\right].
\end{equation}

Furthermore, (\ref{qrfromm}) gives
 \begin{equation}\label{rqproduct}
  rq = 4 \left|m^{(1)}_{12}\right|^2
\end{equation}
Upon evaluation at $x=0$, equations (\ref{qxfromm}), (\ref{qrfromm})-(\ref{rqproduct}) yield
\begin{align}\label{g0g1final}
g_0(t) =& 2im^{(1)}_{12}(t)e^{2i\int^{t}_0 \Delta_2(t')dt'},
		\\ \nonumber
  g_1(t) =& \left(4 m^{(3)}_{12}(t) +  |g_0(t)|^2 m^{(1)}_{12}(t)\right)e^{2i\int^{t}_0 \Delta_2(t')dt'} + ig_0(t) \left(2m^{(2)}_{22}(t) +   |g_0(t)|^2\right),
\end{align}
with
\begin{equation}\label{Delta2final}
\Delta_2(t) = 4\left|m^{(1)}_{12}\right|^4 + 8 \left(\text{Re}\,\left[m^{(1)}_{12} \bar{m}^{(3)}_{12}\right] - \left|m^{(1)}_{12}\right|^2\text{Re}\,\left[m^{(2)}_{22}\right]\right),
 \end{equation}
where $m^{(1)}(t)$, $m^{(2)}(t)$, and $m^{(3)}(t)$ satisfy
\begin{equation}\label{Mtfinal}
M^{(t)}(t, \zeta) = I + \frac{m^{(1)}(t)}{\zeta} + \frac{m^{(2)}(t)}{\zeta^2} + \frac{m^{(3)}(t)}{\zeta^3} + O\left(\frac{1}{\zeta^3}\right), \qquad \zeta \to \infty.
\end{equation}
This establishes equation (\ref{recoverg0g1}).

\subsection{Inverse of spectral map}
It remains to show that the map 
$$\tilde{\Q}:\{A(\zeta), B(\zeta)\} \mapsto \{g_0(t), g_1(t)\}$$
defined in $(v)$ of Proposition \ref{ABprop} is indeed the inverse of the spectral map $\tilde{\mathbb{S}}$. 
In more detail, $\tilde{\Q}$ is defined as follows. 
\begin{enumerate}
\item Given $\{A(\zeta), B(\zeta)\}$, define the jump matrix $J^{(t)}$ by (\ref{Jtdef}). 
\item Find the unique solution $M^{(t)}$ of the RH problem of Proposition \ref{ABprop} with jump matrix $J^{(t)}$. 
\item Compute the functions $m^{(1)}(t)$, $m^{(2)}(t)$, and $m^{(3)}(t)$ from (\ref{Mtfinal}).
\item Compute the function $\Delta_2(t)$ from (\ref{Delta2final}).
\item Obtain $g_0(t)$ and $g_1(t)$ from (\ref{g0g1final}).
\end{enumerate}
As in the $x$-case, we have to show that
$$A_0(\zeta) = A(\zeta), \qquad B_0(\zeta) = B(\zeta),$$
where $\{A_0(\zeta), B_0(\zeta)\}$ is the spectral data corresponding to $\{g_0(t), g_1(t)\}$.
The proof of this statement relies on the dressing method and is analogous to the corresponding proof for the $x$-problem.

\bigskip
\noindent
{\bf Acknowledgement} {\it The author thanks Professor A. S. Fokas for helpful discussions. The research presented in this paper was carried out while the author was supported by a Marie Curie Intra-European Fellowship.}

\bibliographystyle{plain}
\bibliography{is}

\begin{thebibliography}{99}
\small

\bibitem{Mjolhus76}
E. Mjolhus, On the modulational instability of hydromagnetic waves parallel to the magnetic field, 
{\it J. Plasma Phys.} {\bf 16} (1976), 321--334.

\bibitem{Kodama1985}
Y. Kodama, Optical solitons in a monomode fiber, {\it J. Stat. Phys.} {\bf 39} (1985), 597--614.

\bibitem{Agrawal2007}
G. P. Agrawal, {\it Nonlinear fiber optics}, Academic Press, 2007.


\bibitem{K-N}
D. J. Kaup and A. C. Newell, An exact solution for a derivative nonlinear Schr\"odinger equation, {\it J. Math. Phys.} {\bf 19} (1978), 789--801.


\bibitem{K-I}
T. Kawata and H. Inoue, Exact solutions of the derivative nonlinear Schr\"odinger equation under the nonvanishing conditions {\it J. Phys. Soc. Japan} {\bf 44} (1978), 1968--1976. 

\bibitem{IKWS}
Y.-H. Ichikawa, K. Konno, M. Wadati, and H. Sanuki, Spiky soliton in circular polarized Alfv\'en wave,  
{\it J. Phys. Soc. Jpn.} {\bf 48} (1980), 279--286. 


\bibitem{Lashkin}
V. M. Lashkin, $N$-soliton solutions and perturbation theory for the derivative nonlinear Schr\"odinger equation with nonvanishing boundary conditions, 
{\it J. Phys. A} {\bf 40} (2007), 6119--6132.

\bibitem{M-Z}
W. X. Ma and R. Zhou, On inverse recursion operator and tri-Hamiltonian formulation for a Kaup-Newell system of DNLS equations,
{\it J. Phys. A} {\bf 32} (1999), L239--L242.

\bibitem{F1997}
A. S. Fokas, A unified transform method for solving linear and certain nonlinear PDEs, 
{\it Proc. Roy. Soc. Lond.} A {\bf 453} (1997), 1411--1443.

\bibitem{D-Z}
P. A. Deift and X. Zhou, A steepest descent method for oscillatory Riemann-Hilbert problems, 
{\it Ann. Math.} {\bf 137} (1993), 245--338.

\bibitem{F-I-S}
A. S. Fokas, A. R. Its, and L.-Y. Sung, The nonlinear Schr\"odinger equation on the half-line, 
{\it Nonlinearity} {\bf 18} (2005), 1771--1822.

\bibitem{B-F}
J. Bona and A. S. Fokas, Initial-boundary-value problems for linear and integrable nonlinear dispersive PDE's, preprint.

\bibitem{F-I}
A. S. Fokas and A. R. Its, The linearization of the initial-boundary value problem of the nonlinear Schr\"odinger equation,
{\it SIAM J. Math. Anal.} {\bf 27} (1996), 738--764. 

\bibitem{Z-S1}
V. E. Zakharov and A. B. Shabat, A scheme for integrating the nonlinear equations of numerical physics by the method of the inverse scattering problem I, Funct. Anal. Appl. {\bf 8} (1974), 226--235.

\bibitem{Z-S2}
V. E. Zakharov and A. B. Shabat, A scheme for integrating the nonlinear equations of numerical physics by the method of the inverse scattering problem II, Funct. Anal. Appl. {\bf 13} (1979), 166--174.

\bibitem{F2002}
A. S. Fokas, Integrable nonlinear evolution equations on the half-line, 
{\it Comm. Math. Phys.} {\bf 230} (2002), 1--39.


\end{thebibliography}

\end{document}